\documentstyle[12pt]{article}

\def\thebibliography#1{\section*{REFERENCES\markboth
 {REFERENCES}{REFERENCES}}\list
 {[\arabic{enumi}]}{\settowidth\labelwidth{[#1]}\leftmargin\labelwidth
 \advance\leftmargin\labelsep
 \usecounter{enumi}}
 \def\newblock{\hskip .11em plus .33em minus -.07em}
 \sloppy
 \sfcode`\.=1000\relax}

\begin{document}

\begin{titlepage}
\begin{flushright} RAL-95-027\\
\end{flushright}
\vskip 1cm
\begin{center}
{\bf\Large A Nonabelian Yang-Mills Analogue of Classical Electromagnetic
Duality}
\vskip .7cm
{\large Chan Hong-Mo}\\
\vskip .3cm
{\it Rutherford Appleton Laboratory,\\
Chilton, Didcot, Oxon, OX11 0QX, UK.}\\
\vskip .5cm
{\large J. Faridani}\\
\vskip .3cm
{\it Department of Theoretical Physics, Oxford University,\\
1 Keble Road, Oxford, OX1 3NP, UK.}\\
\vskip .5cm
{\large Tsou Sheung Tsun}\\
\vskip .3cm
{\it Mathematical Institute, Oxford University,\\
24-29 St. Giles', Oxford, OX1 3LB, UK.}\\
\end{center}
\begin{abstract}
The classic question of a nonabelian Yang-Mills analogue to electromagnetic
duality is here examined in a minimalist fashion at the strictly 4-dimensional,
classical field and point charge level.  A generalisation of the abelian
Hodge star duality is found which, though not yet known to give dual symmetry,
reproduces analogues to many dual properties of the abelian theory.  For
example, there is a dual potential, but it is a 2-indexed tensor $T_{\mu\nu}$
of the Freedman-Townsend type.  Though not itself functioning as such,
$T_{\mu\nu}$ gives rise to a dual parallel transport, $\tilde{A}_\mu$, for
the phase of the wave function of the colour magnetic charge, this last
being a monopole of the Yang-Mills field but a source of the dual field.
The standard colour (electric) charge itself is found to be a monopole of
$\tilde{A}_\mu$.  At the same time, the gauge symmetry is found doubled from
say $SU(N)$ to $SU(N) \times SU(N)$.  A novel feature is that all equations of
motion, including the standard Yang-Mills and Wong equations, are here
derived from a `universal' principle, namely the Wu-Yang (1976) criterion
for monopoles, where interactions arise purely as a consequence of the
topological definition of the monopole charge.  The technique used is the
loop space formulation of Polyakov (1980).
\end{abstract}
\end{titlepage}

\section{Introduction}
Recently, questions of electromagnetic duality have again come to the fore.
Apart from the related duality in superstring theory
\cite{strom}-\cite{gauntlett},
there have been much interest in dual symmetric
actions for field theories \cite{schwarz} and of course the recent work on
electromagnetic duality in supersymmetric theories of Seiberg, Witten and
co-workers \cite{seibwitt}-\cite{wittenmon}.
Although these recent works are based
on sophisticated solitonic structures in sophisticated theories, they
originate nevertheless from the simple property of the Maxwell theory of
being invariant under the interchange of electricity and magnetism, which
though long known, has retained a continued and recurrent interest among
physicists.

The reason is that dual symmetry in electromagnetism has associated with it
many
important consequences.  For example, the fact that the Maxwell field
$F_{\mu\nu}$ is a gauge field derivable from a gauge potential $A_\mu$
implies that the dual field ${\mbox{\mbox{}$^*\!$}} F_{\mu\nu}$
must also be derivable
from some potential ${\tilde A}_\mu$.  And just as $A_\mu$ acts as the
parallel phase transport for the wave functions of electric charges, so its
dual partner $\tilde{A}_\mu$ will act as the parallel phase transport for
the wave functions of particles carrying a magnetic charge. Hence, since the
theory is invariant under a $U(1)$ gauge transformation: $A_\mu
\longrightarrow A_\mu + \partial_\mu \Lambda$, it follows that it must
also be invariant under a ${\tilde U}(1)$ transformation: ${\tilde A}_\mu
\longrightarrow {\tilde A}_\mu + \partial_\mu {\tilde \Lambda}$, giving
in all a $U(1) \times {\tilde U}(1)$ invariance.  Further, although it is
conventional to regard an electric charge as a source and a magnetic charge
as a monopole of the Maxwell field $F_{\mu\nu}$, it is equally admissible,
because of dual symmetry, to consider an electric charge as a monopole and a
magnetic charge as a source of the dual Maxwell field
$\mbox{\mbox{}$^*\!$} F_{\mu\nu}$.
\footnote{Here, a source is considered to give rise to a current represented
by the divergence of the field, while a monopole results from a topologically
nontrivial configuration of the potential.}  Thus, it is possible to apply the
Wu-Yang criterion \cite{wuyang} for monopoles to derive, for example, the
Lorentz equation for either an electric or a magnetic charge from the
monopole's definition as a topological obstruction without explicitly
introducing an interaction term into the action.  Indeed, in Charts I and
II below, we have listed a whole complex of properties of electromagnetism,
some perhaps less well-known than the above-mentioned, which are all
connected in some way with the concept of duality.

Given the importance of nonabelian Yang-Mills theories in present day physics,
it is natural to ask whether the notion of electromagnetic duality extends
to them as well.  In the recent developments, this question  is addressed in a
broad daring fashion by Seiberg, Witten and co-workers, in which
electromagnetic
duality is embedded in a grand unified (at present supersymmetric) quantum
field framework, where charges (whether electric or magnetic) appear as
't Hooft-Polyakov solitons.  They not only give strong evidence for an
exact duality in supersymmetric Yang-Mills theories as suggested by Olive
and Montonen \cite{olivemont} but trigger also a breakthrough in 4-manifold
topology. (See \cite{wittenmon} and recent work of C. Taubes on Donaldson
invariants.) More work has also been done to support and generalise the above
\cite{aharony} The result is an imposing structure with rich and exciting
properties which is being diligently explored by both the physics and
mathematics communities.

Our present work addresses the same basic question but starts from a
complementary minimalist standpoint.  The idea is that since the
attractive dual properties listed above occur for electromagnetism already
at the classical field level with point charges, it makes sense also to ask
whether (nonabelian) Yang-Mills theories may share some of these properties
already at the same classical field and point charge level.

In terms of the Maxwell fields $F_{\mu\nu}$, symmetry
under the interchange of electricity and magnetism, i.e. $\bf{E} \rightarrow
-\bf{H}, \bf{H} \rightarrow \bf{E}$, means an invariance under the Hodge
star operation: $F_{\mu\nu} \rightarrow \mbox{\mbox{}$^*\!$} F_{\mu\nu}
= -\mbox{\small $\frac{1}{2}$} \epsilon_{\mu\nu\rho\sigma} F^{\rho\sigma}$.  In
Yang-Mills
theory, on the other hand, although the action is still invariant (apart from
a sign) under $F_{\mu\nu} \rightarrow \mbox{\mbox{}$^*\!$}
 F_{\mu\nu}$, the dual field
$\mbox{\mbox{}$^*\!$}
 F_{\mu\nu}$, in contrast to $F_{\mu\nu}$, is not in general
a gauge field derivable from a potential. \cite{guyang}  Hence, if duality
is still taken to mean the Hodge star operation $F_{\mu\nu} \rightarrow
\mbox{\mbox{}$^*\!$}
 F_{\mu\nu}$ for the nonabelian theory, then strict dual symmetry
is lost. However, it is possible to entertain the notion of a generalised
duality relationship in Yang-Mills theory, which, though reducing to to the
Hodge star operation in the abelian theory, is not identical to it when the
theory is nonabelian.  The only question is whether such a notion would be
profitable.  Considerations can then be taken at two levels.  One notes
that the `dual' properties of electromagnetism listed in Charts I and II
do not by themselves imply dual symmetry although they are consequences of
it.  Thus a generalised duality relationship may be such as to reproduce all
the dual properties of Charts I and II for Yang-Mills fields without restoring
dual symmetry and it would still be a very useful concept.

What we wish to exhibit below is a generalised duality relationship in
Yang-Mills fields, which though not known as yet to give a fully-fledged dual
symmetry manages nevertheless to reproduce nonabelian analogues to all
the dual properties of electromagnetism listed in Charts I and II.  For
example, one finds that there is in nonabelian theory indeed a
`generalised dual potential', whose `curl' or `exterior derivative', though
not $\mbox{\mbox{}$^*\!$}
 F_{\mu\nu}$, is a quantity closely related to it.  This
`dual potential' is not a vector quantity but an antisymmetric second-rank
tensor $T_{\mu\nu}$ of the type first found in superstring
theory \cite{tensorss}, but although it does not itself act as the
parallel phase transport for wave functions of the colour magnetic
charge, it gives rise to another quantity $\tilde{A}_\mu$ which plays
that role.  In terms of $T_{\mu\nu}$, the Yang-Mills action takes on a
Freedman-Townsend form \cite{freedtown} invariant under the transformation
$T_{\mu\nu} \longrightarrow T_{\mu\nu} +\delta_{[\mu} \Lambda_{\nu]}$, which
induces in turn in $\tilde{A}_\mu$ a new $SU(N)$ transformation of odd parity,
giving thus, as in the abelian theory, a doubling of the original gauge
symmetry, from $SU(N)$ to $SU(N) \times \widetilde{SU(N)}$.
Further, it will appear that the sources of one potential can again be regarded
as monopoles of its `dual' and vice versa, so that, if we call a source of
the Yang-Mills potential $A_\mu$ a `colour electric' and a monopole of
$A_\mu$ a `colour magnetic' charge, then a `colour electric' charge can
also be regarded as a `monopole' and a `colour magnetic' charge a source
of the `dual potential' $T_{\mu\nu}$.  It even follows then that both the
Yang-Mills equation and its classical limit, the Wong equation \cite{wong},
can be derived from the definition of a monopole using the above-mentioned
Wu-Yang criterion.  The full result is summarised in Charts III and IV below,
which are seen to parallel closely Charts I and II of the abelian theory.

We note that in spite of this close analogy to the abelian theory, the fact
remains that the nonabelian theory is not yet known to be fully symmetric.
Thus although both `colour electric' and `colour magnetic' charges can be
considered as monopoles in some field so that the Wu-Yang criterion can be
used in each case to derive the equations of motion, the two `dual' sets of
equations so obtained need no longer describe the same dynamics as they did
in the abelian theory.  For the `colour electric' charge, the equations
turn out to be exactly those of the standard Yang-Mills theory, in spite of
their very different derivation from the conventional.  The equations for
the `colour magnetic' charge, on the other hand, may describe a different
dynamics which is only just beginning to be explored. \cite{jackie}

In exhibiting the analogue duality of Yang-Mills theory, we have used the
loop space description of gauge theory in terms of some loop variables first
introduced by Polyakov in 1980, \cite{polyakov}.  Although we are unsure
whether
this apparatus is essential, we believe it to be appropriate for the
following reason.  The question of duality is closely connected to the
existence or otherwise of a `dual potential', which is in turn intimately
related to the presence or absence of monopoles.  The virtue of the
Polyakov loop variables is that, unlike the standard local variables
$A_\mu$ and $F_{\mu\nu}$, they remain patch-independent in the Wu-Yang
\cite{patching} sense (or nonsingular in the sense of not having any Dirac
strings) even in the presence of monopoles.  They are thus particularly
suited to the problem.

It should be stressed, however, that in spite of this loop formulation, there
has been no input other than the standard Yang-Mills theory in deriving the
results, and that we have worked throughout in a space-time of strictly
4 dimensions.  The dual properties we have found are thus of ordinary
Yang-Mills fields, only cast in a somewhat unconventional language.

In order to exhibit fully the nonabelian dual structure under consideration,
much effort in this paper has to be spent in reformulating and reorganising
information of results contained in our earlier work.  The main new specific
results of this paper are Sections 4 and 6, of which the most interesting
we think is in Section 6 where it is shown that ordinary colour (electric)
charges are monopoles of the dual formulation and that their known dynamics
can be derived from the Wu-Yang criterion for monopoles.

\setcounter{equation}{0}
\section{Duality in Electromagnetism}

Let us begin by recalling some of the known dual properties of
electromagnetism so as to facilitate comparison with the new results we
shall later derive for nonabelian theories. \cite{annphys,book,physrev}

Consider first pure electrodynamics, with neither electric
nor magnetic charges in the theory.  One starts then usually with a gauge
field $F_{\mu\nu}(x)$ derivable from a potential $A_\mu(x)$, thus:
\begin{equation}
F_{\mu\nu}(x) = \partial_\nu A_\mu(x) - \partial_\mu A_\nu(x).
\label{fmunu}
\end{equation}
The field action is taken to be:\footnote{In our convention, $g_{\mu\nu}
= (+,-,-,-), \epsilon_{0123} = 1.$}
\begin{equation}
{\cal A}^0_F = -\frac{1}{16 \pi} \int d^4x F_{\mu\nu}(x) F^{\mu\nu}(x),
\label{scriptA0F}
\end{equation}
where $F_{\mu\nu}(x)$ is given in terms of $A_\mu(x)$ through (\ref{fmunu}),
and equations of motion are to be obtained by extremising ${\cal A}^0_F$
with respect to $A_\mu(x)$, giving:
\begin{equation}
\partial^\nu F_{\mu\nu}(x) = 0.
\label{maxwelleq}
\end{equation}
Define next the `dual field' $\mbox{\mbox{}$^*\!$} F_{\mu\nu}(x)$ as:
\begin{equation}
\mbox{\mbox{}$^*\!$}
 F_{\mu\nu}(x) = -\mbox{\small $\frac{1}{2}$} \epsilon_{\mu\nu\rho\sigma}
F^{\rho\sigma}(x),
\label{fstar}
\end{equation}
where the Hodge star operation * exchanges electricity and magnetism thus:
$\bf{E} \rightarrow -\bf{H},\bf{H} \rightarrow \bf{E}$.
In terms of $\mbox{\mbox{}$^*\!$} F_{\mu\nu}(x)$, the equation
(\ref{maxwelleq}) is equivalent
to the Bianchi identity:
\begin{equation}
\partial_\lambda \mbox{\mbox{}$^*\!$} F_{\mu\nu}(x) + \partial_\mu
\mbox{\mbox{}$^*\!$} F_{\nu\lambda}(x)
   + \partial_\nu \mbox{\mbox{}$^*\!$} F_{\lambda\mu}(x) = 0,
\label{dbianchi}
\end{equation}
which implies, in this abelian case, via the Poincar\'e lemma in 4-dimensional
space-time that $\mbox{\mbox{}$^*\!$} F_{\mu\nu}(x)$ must also be a gauge field
locally
derivable from some potential ${\tilde A}_\mu(x)$, thus:
\begin{equation}
\mbox{\mbox{}$^*\!$} F_{\mu\nu}(x) = \partial_\nu {\tilde A}_\mu(x) -
\partial_\mu
   {\tilde A}_\nu(x).
\label{dfmunu}
\end{equation}
Hence one sees that starting with a gauge field $F_{\mu\nu}(x)$ satisfying
(\ref{fmunu}), one ends up via the the equation of motion (\ref{maxwelleq})
with an $\mbox{\mbox{}$^*\!$} F_{\mu\nu}(x)$ which is also a gauge field
derivable from
a potential.  This is an essential feature of electromagnetic duality.

Further, it follows that in addition to the original gauge invariance under
the transformation:
\begin{equation}
A_\mu(x) \longrightarrow A_\mu(x) +\partial_\mu \Lambda(x),
\label{utransf}
\end{equation}
there is a gauge invariance also under the transformation:
\begin{equation}
{\tilde A}_\mu(x) \longrightarrow {\tilde A}_\mu(x) + \partial_\mu
   {\tilde \Lambda}(x),
\label{uttransf}
\end{equation}
giving the theory in all a $U(1) \times U(1)$ gauge symmetry.  The second
$U(1)$ has in fact the opposite parity to the first because of the Hodge
star operation and will henceforth be referred to as the
${\tilde U}$-invariance to distinguish it from the first $U(1)$.  Note,
however, that since $\tilde{A}_\mu(x)$ is related to $A_\mu(x)$ through
(\ref{fmunu}), (\ref{fstar}), and (\ref{dfmunu}), the two $U(1)$ symmetries
do not correspond to two distinct physical degrees of freedom.

There is an alternative method for treating the above problem which, though
yielding nothing new in pure electrodynamics, is particularly amenable for
future adaptation both to electrodynamics with electric or magnetic charges
and to nonabelian Yang-Mills theories. \cite{annphys,book,physrev} Instead
of the potential $A_\mu(x)$, one adopts $F_{\mu\nu}(x)$ as field variables,
but under the
constraint:
\begin{equation}
\partial_\lambda F_{\mu\nu}(x) + \partial_\mu F_{\nu\lambda}(x)
   + \partial_\nu F_{\lambda\mu}(x) = 0,
\label{bianchi}
\end{equation}
or equivalently:
\begin{equation}
\partial^\nu \mbox{\mbox{}$^*\!$} F_{\mu\nu}(x) = 0.
\label{gausslaw}
\end{equation}
The set of variables $\{F_{\mu\nu}(x)\}$ is of course inherently redundant,
comprising more components than the original set $\{A_\mu(x)\}$ which we
know is already sufficient for describing our system.  However, by imposing
the constraint (\ref{bianchi}) or (\ref{gausslaw}), this redundancy is removed
since (\ref{bianchi}) is the Bianchi identity for $F_{\mu\nu}(x)$ and
hence, as above for the case of $\mbox{\mbox{}$^*\!$} F_{\mu\nu}(x)$, ensures
via the
Poincar\'e lemma that $F_{\mu\nu}(x)$ is a gauge field derivable locally
from a potential as in (\ref{fmunu}).  Hence the problem as formulated
now in terms of $F_{\mu\nu}(x)$ is the same as the original problem
formulated in terms of $A_\mu(x)$.  Equations of motion are now to be
obtained by extremising ${\cal A}^0_F$ in (\ref{scriptA0F}) with respect
to $F_{\mu\nu}(x)$ but under the constraint (\ref{gausslaw}).  Introducing
then Lagrange multipliers $\lambda_\mu(x)$ and forming the action:
\begin{equation}
{\cal A}_F = -\frac{1}{16 \pi} \int d^4x F_{\mu\nu}(x) F^{\mu\nu}(x)
   + \int d^4x \lambda_\mu(x) \partial_\nu \mbox{\mbox{}$^*\!$} F^{\mu\nu}(x),
\label{scriptAF}
\end{equation}
we then obtain the equation:
\begin{equation}
F_{\mu\nu}(x) = 4 \pi \{\mbox{\small $\frac{1}{2}$}
\epsilon_{\mu\nu\rho\sigma}[\partial^\sigma
   \lambda^\rho(x) - \partial^\rho \lambda^\sigma(x)]\},
\label{finlambda}
\end{equation}
or:
\begin{equation}
\mbox{\mbox{}$^*\!$} F_{\mu\nu}(x) = 4 \pi [\partial_\nu \lambda_\mu(x) -
\partial_\mu
   \lambda_\nu(x)],
\label{fsinlambda}
\end{equation}
which states that $\mbox{\mbox{}$^*\!$} F_{\mu\nu}(x)$ is a gauge field with $4
\pi$ times
the Lagrange multiplier $\lambda_\mu(x)$ as the gauge potential:
\begin{equation}
{\tilde A}_\mu(x) = 4 \pi \lambda_\mu(x).
\label{atinlambda}
\end{equation}
In turn, (\ref{fsinlambda}) implies that $\mbox{\mbox{}$^*\!$} F_{\mu\nu}(x)$
must satisfy
the Bianchi identity (\ref{dbianchi}), or equivalently the Maxwell equation
(\ref{maxwelleq}).  In other words, one obtains just exactly the same result as
before, as expected.  However, this alternative treatment has two new
features, which will facilitate the analysis of the duality question.
First, it is formulated in terms of gauge invariant variables
$F_{\mu\nu}(x)$ which, as we shall see later, are particularly useful in
solving the parallel problem in the presence of monopoles.  Secondly, the dual
potential appears automatically as the Lagrange multiplier to the constraint
imposed on these variables to remove their inherent redundancy, which will
afford later a valuable hint on how the dual potential may be generalised to
the nonabelian theory.

{}From (\ref{fstar}), one sees that:
\begin{equation}
\mbox{\mbox{}$^*\!$} (\mbox{\mbox{}$^*\!$} F_{\mu\nu}(x)) = - F_{\mu\nu}(x),
\label{fstarstar}
\end{equation}
meaning that (apart from a sign) $F_{\mu\nu}(x)$ is also the dual of
$\mbox{\mbox{}$^*\!$} F_{\mu\nu}(x)$, and that when expressed in terms of
$\mbox{\mbox{}$^*\!$} F_{\mu\nu}(x)$, ${\cal A}^0_F$ in (\ref{scriptA0F}) is of
exactly
the same form (also apart from a sign):
\begin{equation}
{\tilde {\cal A}}^0_F = \frac{1}{16 \pi} \int d^4x \mbox{\mbox{}$^*\!$}
F_{\mu\nu}(x)
   \mbox{\mbox{}$^*\!$} F^{\mu\nu}(x).
\label{scrptAt0F}
\end{equation}
{}From these facts, it follows that all our preceding arguments would remain
valid under the interchange of $F_{\mu\nu}(x)$ with $\mbox{\mbox{}$^*\!$}
F_{\mu\nu}(x)$
and $A_\mu(x)$ with ${\tilde A}_\mu(x)$, which is what one usually calls
dual symmetry.

For easy reference in the future, the dual structure of pure electrodynamics
as elucidated above is summarised in the flowchart Chart I.

Next, consider electrodynamics in the presence of electric or magnetic
charges.  Conventionally, one is used to regarding electric charges as
`sources'
and magnetic charges as `monopoles' of the Maxwell field $F_{\mu\nu}(x)$ in
the sense that electric charges give nonzero divergences to the field while
magnetic charges result from nontrivial topological configurations of the
potential. Because of the dual symmetry described above, however, it is equally
valid
to regard electric charges as `monopoles' and magnetic charges as `sources'
of the dual Maxwell field $\mbox{\mbox{}$^*\!$} F_{\mu\nu}(x)$.

In formulating the interaction of an electric or a magnetic charge with
the electromagnetic field, the usual approach is to regard the charge
as a source. Thus, for a classical point electric charge $e$, one
chooses usually to work with the Maxwell field $F_{\mu\nu}(x)$ in which
$e$ appears as a source, and write for the action:
\begin{equation}
{\cal A}^0 = -\frac{1}{16 \pi} \int d^4x F_{\mu\nu}(x) F^{\mu\nu}(x)
   - m \int d\tau.
\label{scriptA0c}
\end{equation}
Or else, if the charge $e$ is to be carried by a quantum particle described
by, say, a Dirac wave function $\psi(x)$, we write:
\begin{equation}
{\cal A}^0 = -\frac{1}{16 \pi} \int d^4x F_{\mu\nu}(x) F^{\mu\nu}(x)
   +\int d^4x {\bar \psi}(x) (i \partial_\mu \gamma^\mu - m) \psi(x).
\label{scriptA0q}
\end{equation}
The interaction between the charge and the field is then introduced as an
additional term in the action, thus:
\begin{equation}
{\cal A}_I = e \int A_\mu(x) dx^\mu,
\label{scriptAIc}
\end{equation}
for the classical charge, where the integral is to be taken along the
world-line $Y^\mu(\tau)$ of the particle, or else:
\begin{equation}
{\cal A}_I = e \int d^4x {\bar \psi}(x) A_\mu(x) \gamma^\mu \psi(x),
\label{scriptAIq}
\end{equation}
for the quantum charge.

The alternative approach given above for pure electromagnetism was originally
formulated for charges treated as monopoles rather than as sources, in which
case their interaction appears as a unique consequence of the topological
definition of a monopole.  This is contained in a paper of Wu and Yang in
1976 and will be referred to here henceforth as the Wu-Yang criterion.
\cite{wuyang,annphys,book,physrev}  Intuitively, the idea is as follows.
Monopoles are topological obstructions in gauge fields, because of which
their charges are necessarily quantised and conserved.  That there exists
a monopole at some point in space means that the gauge field has a certain
topological configuration around that point.  Thus, if the monopole moves
to another position, the gauge field will have to rearrange itself so
as to maintain the same topological configuration around the new point.
Hence, the definition of a monopole as a topological obstruction
already contains within it an implied coupling between the monopole's position
and the surrounding field, or in physical terms a charge-field interaction.
The only question is whether this intrinsic interaction is the same as that
of the usual formulation.

To answer this question, we first recall briefly the description of a monopole
as a topological obstruction. \cite{book}  Since electrodynamics is
dual symmetric, we can work either with an electric charge regarded as a
monopole of $\mbox{\mbox{}$^*\!$} F_{\mu\nu}(x)$ or with a magnetic charge
regarded as
a monopole of $F_{\mu\nu}(x)$.  We choose here to illustrate with the latter
since the notation there is more familiar.  Take a one-parameter family
of closed loops in space enveloping a closed surface $\Sigma$.  Label the
loops as $\{C_t\}$ with the parameter $t$ ranging in value from $0$ to
$2 \pi$ so that at $t = 0$ and $t = 2 \pi$ $C_t$ shrinks to the point $P_0$.
For each of these loops, one can define a phase factor:
\begin{equation}
\Phi(C) = \exp ie \oint_C A_\mu(x) dx^\mu,
\label{PhiofC}
\end{equation}
which is an element of the gauge group $U(1)$.  As $t$ varies from $0$
to $2 \pi$, the image of $C_t$ by $\Phi$ in $U(1)$ will trace a closed
curve, say $\Gamma_\Sigma$, in $U(1)$, beginning and ending at the group
identity.  The gauge group $U(1)$ here has the topology of a circle $S^1$
so that the closed curve $\Gamma_\Sigma$ must wind around this circle an
integral number of times.  In other words, the total change in phase of
$\Phi(C_t)$ for $t = 0 \rightarrow 2 \pi$ must equal $2 \pi n$, with $n$
integral.  This value of $n$, being discrete, cannot be changed by any
continuous deformation of the surface $\Sigma$.  Hence, for $n \neq 0$,
the gauge field will behave as if there is some object enclosed inside
$\Sigma$ which forbids $\Sigma$ from trivially shrinking to zero.  This is
what is termed a topological obstruction.  Further, it is easily seen that
the total change in phase of $\Phi(C_t)$ for $t = 0 \rightarrow 2 \pi$
is just the total magnetic flux flowing out of the surface $\Sigma$, or else
$4 \pi$ times the magnetic charge enclosed inside $\Sigma$.  Hence, we conclude
that the topological obstruction here represents a magnetic monopole with
a quantised charge ${\tilde e}$.

To derive the equations of motion of a classical magnetic charge interacting
with an electromagnetic field, our considerations above require that we
extremise the action (\ref{scriptA0c}) subject to the condition that the
particle $Y^\mu(\tau)$ represents a topological obstruction of the type
described in the preceding paragraph. It can readily be shown that such a
condition can be written in the following Lorentz covariant form:
\begin{equation}
\partial_\nu \mbox{\mbox{}$^*\!$} F^{\mu\nu}(x) = -4 \pi {\tilde e} \int d\tau
   \frac{d Y^\mu(\tau)}{d\tau} \delta(x - Y(\tau)).
\label{gausslawc}
\end{equation}
In solving the variational problem for the equations of motion, one can in
principle retain the gauge potential $A_\mu(x)$ as variables but since, as
noted above, $A_\mu(x)$ has to be patched in the presence of the monopole,
this would make the problem rather unwieldy, especially since the patches
will have to depend on the position $Y$ of the monopole.  This suggests
that one adopts instead the alternative approach using $F_{\mu\nu}(x)$ as
variables but, in that case, one would have to constrain $F_{\mu\nu}(x)$
so as to ensure that their inherent redundancy be removed or that the
potential $A_\mu(x)$ can be recovered from them.  The beauty, however, is
that the constraint we need for removing the redundancy is already exactly
contained in the condition (\ref{gausslawc}) we wish to impose.  Indeed,
(\ref{gausslawc}) implies that, except on the monopole world-line
$Y^\mu(\tau)$,
the Bianchi identity (\ref{gausslaw}) or (\ref{bianchi}) is satisfied, so
that by the Poincar\'e lemma $F_{\mu\nu}(x)$ is derivable from
a potential, whereas on $Y^\mu(\tau)$ itself, we do not expect to have a
potential $A_\mu(x)$ in any case.  Thus, in analogy to the previous treatment
of pure electrodynamics, one can just use $F_{\mu\nu}(x)$ as field variables
in place of $A_\mu(x)$ so long as (\ref{gausslawc}) is satisfied, and so avoid
all problems with patching.

To derive the equations of motion then, we extremise the free action
(\ref{scriptA0c}) with respect to $F_{\mu\nu}(x)$ and $Y^\mu(\tau)$ under
the constraint (\ref{gausslawc}).  Introducing Lagrange multipliers
$\lambda_\mu(x)$ as before, we extremise instead:
\begin{equation}
{\cal A} = {\cal A}^0 + \int d^4x \lambda_\mu(x) \{\partial_\nu
\mbox{\mbox{}$^*\!$}
   F^{\mu\nu}(x) + 4 \pi {\tilde e} \int d\tau \frac{d Y^\mu(\tau)}{d \tau}
   \delta (x - Y(\tau))\}.
\label{scriptAc}
\end{equation}
The solution is straightforward, giving again (\ref{fsinlambda}) as well as:
\begin{equation}
m \frac{d^2Y_\mu(\tau)}{d \tau^2} = -4 \pi {\tilde e} \{\partial_\nu
   \lambda_\mu(Y(\tau)) - \partial_\mu \lambda_\nu(Y(\tau))\}
   \frac{dY^\nu(\tau)}{d \tau},
\label{eleqc}
\end{equation}
which, on eliminating $\lambda_\mu(x)$ by (\ref{fsinlambda}), yields
respectively (\ref{maxwelleq}) and:
\begin{equation}
m \frac{d^2Y^\mu(\tau)}{d \tau^2} = - {\tilde e} \mbox{\mbox{}$^*\!$}
F^{\mu\nu}(Y(\tau))
   \frac{dY_\nu(\tau)}{d \tau}.
\label{duallorentz}
\end{equation}
Together with the original constraint (\ref{gausslawc}), (\ref{maxwelleq}) and
(\ref{duallorentz}) then constitute the complete set of equations of motion
for the magnetic charge interacting with the electromagnetic field.

Since the system is dual symmetric, exactly the same arguments would apply
for an electric charge when considered as a monopole of the dual Maxwell
field $\mbox{\mbox{}$^*\!$} F_{\mu\nu}(x)$.  The constraint defining the charge
as a
topological obstruction would then be:
\begin{equation}
\partial_\nu F^{\mu\nu}(x) = -4 \pi e \int d\tau \frac{dY^\mu(\tau)}{d\tau}
   \delta (x - Y(\tau)),
\label{maxwelleqc}
\end{equation}
under which the action (\ref{scriptA0c}), or equivalently the action in terms
of $\mbox{\mbox{}$^*\!$} F_{\mu\nu}(x)$:
\begin{equation}
{\tilde A}^0 = \frac{1}{16 \pi} \int d^4x \mbox{\mbox{}$^*\!$} F_{\mu\nu}(x)
   \mbox{\mbox{}$^*\!$} F^{\mu\nu}(x) - m \int d\tau,
\label{scriptAt0c}
\end{equation}
has to be extremised with respect to $\mbox{\mbox{}$^*\!$} F_{\mu\nu}(x)$ and
$Y^\mu(\tau)$
as variables.  Proceeding as above, we get:
\begin{equation}
\mbox{\mbox{}$^*\!$} F_{\mu\nu}(x) = 4 \pi \{\mbox{\small $\frac{1}{2}$}
\epsilon_{\mu\nu\rho\sigma}
   [\partial^\sigma {\tilde \lambda}^\rho(x) - \partial^\rho
   {\tilde \lambda}^\sigma(x)]\},
\label{fsinlambdat}
\end{equation}
and:
\begin{equation}
m \frac{d^2Y^\mu(\tau)}{d\tau^2} = - e F^{\mu\nu}(Y(\tau)) \frac{dY_\nu(\tau)}
   {d\tau}.
\label{lorentz}
\end{equation}
(\ref{fsinlambdat}) says that $F_{\mu\nu}(x)$ can be derived from a
potential as in (\ref{fmunu}) with:
\begin{equation}
A_\mu(x) = 4 \pi {\tilde \lambda}_\mu(x),
\label{ainlambdat}
\end{equation}
The equations (\ref{maxwelleqc}), (\ref{fmunu}), and (\ref{lorentz}) are
exactly the standard Maxwell and Lorentz equations governing the motion of
an electric charge moving in an electromagnetic field as normally derived
by extremising:
\begin{equation}
{\cal A}' = {\cal A}^0 + {\cal A}_I
\label{scriptAp}
\end{equation}
with respect to $A_\mu(x)$ and $Y^\mu(\tau)$ for ${\cal A}^0$ in
(\ref{scriptA0c}) and ${\cal A}_I$ in (\ref{scriptAIc}).  The present approach
is thus entirely equivalent to the conventional approach as expected, but has
the virtue of giving the interaction uniquely as a consequence of the
topological definition of the charge without introducing an interaction term
in the action, the choice of which in the conventional approach seems to
admit additional freedom.

The formulation for a quantum charged particle when considered as a monopole
is similar.  For a magnetic charge, we extremise then ${\cal A}^0$ in
(\ref{scriptA0q}) with respect to $F_{\mu\nu}(x)$ and $\psi(x)$ under the
constraint:
\begin{equation}
\partial_\nu \mbox{\mbox{}$^*\!$} F^{\mu\nu}(x) = - 4 \pi {\tilde e} {\bar
\psi}(x)
   \gamma^\mu \psi(x),
\label{gausslawq}
\end{equation}
where we have just replaced in (\ref{gausslawc}) the classical current by
the quantum mechanical current.  Using again the method of Lagrange
multipliers and varying:
\begin{equation}
{\cal A} = {\cal A}_0 + \int d^4x \lambda_\mu(x) \{\partial_\nu
   \mbox{\mbox{}$^*\!$} F^{\mu\nu}(x) + 4\pi \tilde{e} \bar{\psi}(x) \gamma^\mu
\psi(x) \},
\label{scriptAq}
\end{equation}
with respect to $F_{\mu\nu}(x)$ and $\psi(x)$ we
obtain (\ref{finlambda}) and:
\begin{equation}
(i \partial_\mu \gamma^\mu - m) \psi(x) = - {\tilde e} {\tilde A}_\mu(x)
   \gamma^\mu \psi(x),
\label{ddiraceq}
\end{equation}
for ${\tilde A}_\mu(x)$ as given in (\ref{atinlambda}), which are exactly the
dual of the standard equations for an electric charge moving in a Maxwell
field.
Indeed, because of dual symmetry, one readily sees that had we started with
an electric charge considered as a monopole in the dual Maxwell field
$\mbox{\mbox{}$^*\!$} F_{\mu\nu}(x)$, we would have obtained the standard
equations for
an electric charge.

Finally, we note that as for pure electrodynamics, the systems with electric
or magnetic charges will still have the gauge symmetry doubled to
$U(1) \times \widetilde{U(1)}$.  Take the action (\ref{scriptAq}) for example,
which will be useful later for comparison with the parallel in the nonabelian
theory. Under the original $U(1)$ symmetry, all the quantities appearing in
${\cal A}$ are invariant, including $\psi(x)$ which, though magnetically
charged, is electrically neutral.  Under a ${\tilde U}$-transformation, on
the other hand, $\mbox{\mbox{}$^*\!$} F_{\mu\nu}(x)$ is invariant,
$\lambda_\mu(x)$
transforms as in (\ref{uttransf}) and $\psi(x)$ transforms as:
\begin{equation}
\psi(x) \longrightarrow  (1 + i{\tilde e} {\tilde \Lambda}(x)) \psi(x).
\label{uttransfpsi}
\end{equation}
Hence, in (\ref{scriptAq}), ${\cal A}^0_F$ in ${\cal A}^0$ is invariant,
and so is the first term in the integral as can be seen by an integration
by parts using the antisymmetry of $\mbox{\mbox{}$^*\!$} F_{\mu\nu}(x)$, while
the
variation from the second term in the integral will cancel with the
increment from the free action of $\psi$ in (\ref{scriptA0q}).  Thus the
symmetry is indeed doubled as claimed.

For easier comparison with later extensions to nonabelian theories, the dual
structure as revealed in the above treatment is summarised in Chart II.

\setcounter{equation}{0}
\section{Pure Yang-Mills Theory - Direct Formulation}

In Yang-Mills theory, one usually starts also with a gauge field
$F_{\mu\nu}(x)$
derivable from a potential, thus:\footnote{Although our results apply to all
$su(N)$ theories, they will be given explicitly only for $su(2)$ for
convenience.  In our convention for $su(2)$, $B = B^it_i, t_i = \tau_i/2,
\,\mbox{\rm Tr} B = 2 \times$ sum of diagonal elements, so that $\,\mbox{\rm
Tr} (t_i t_j)
= \delta_{ij}$.}
\begin{equation}
F_{\mu\nu}(x) = \partial_\nu A_\mu(x) - \partial_\mu A_\nu(x)
   +ig [A_\mu(x), A_\nu(x)].
\label{Fmunu}
\end{equation}
Equations of motion are obtained by extremising the action:
\begin{equation}
{\cal A}^0_F = -\frac{1}{16 \pi} \int d^4x \,\mbox{\rm Tr} \{F_{\mu\nu}(x)
F^{\mu\nu}(x)\},
\label{ScriptA0F}
\end{equation}
with respect to $A_\mu(x)$.  This gives:
\begin{equation}
D_\nu F^{\mu\nu}(x) = 0,
\label{yangmillseq}
\end{equation}
where $D_\nu$ is the covariant derivative defined as:
\begin{equation}
D_\mu = \partial_\mu -ig [A_\mu(x), \ \ \ ].
\label{covderiv}
\end{equation}
So far, the analogy to electromagnetism is still rather close.

However, the result in electromagnetism that the equation of motion
(\ref{maxwelleq}) implies the existence of a potential also for
$\mbox{\mbox{}$^*\!$} F_{\mu\nu}(x)$ no longer holds in general for nonabelian
Yang-Mills
theories. \cite{guyang}  One may still define, of course, a dual field
$\mbox{\mbox{}$^*\!$} F_{\mu\nu}(x)$ as in (\ref{fstar}) for which the equation
(\ref{yangmillseq}) appears again as the Bianchi identity:
\begin{equation}
D_\lambda \mbox{\mbox{}$^*\!$} F_{\mu\nu}(x) + D_\mu \mbox{\mbox{}$^*\!$}
F_{\nu\lambda}(x)
   + D_\nu \mbox{\mbox{}$^*\!$} F_{\lambda\mu}(x) = 0.
\label{Bianchi}
\end{equation}
However, in the absence of a nonabelian analogue to the Poincar\'e lemma,
this no longer implies the existence of a `dual potential' related
to $\mbox{\mbox{}$^*\!$} F_{\mu\nu}(x)$ in the same way that $A_\mu(x)$ is
related to
$F_{\mu\nu}(x)$ in (\ref{Fmunu}).  Indeed, it has been shown explicitly by
Gu and Yang \cite{guyang} that for some particular examples of
$\mbox{\mbox{}$^*\!$} F_{\mu\nu}(x)$ satisfying (\ref{Bianchi}) there does not
exist
any ${\bar A}_\mu(x)$ such that:
\begin{equation}
\mbox{\mbox{}$^*\!$} F_{\mu\nu}(x) = \partial_\nu {\bar A}_\mu(x) -\partial_\mu
   {\bar A}_\nu(x) +ig [{\bar A}_\mu(x), {\bar A}_\nu(x)].
\label{Abar}
\end{equation}

The failure of (\ref{Abar}) destroys any hope of a dual symmetry for nonabelian
Yang-Mills theories in strict analogy to that in electromagnetism.  However,
there may be a possibility that by defining a different dual relationship for
Yang-Mills fields which, though reducing to ordinary electromagnetic duality
for the abelian theory, yet allows for the existence of a generalised dual
potential.  The only question is how many of the so-called `dual properties'
of electromagnetism are retained in this generalised dual relationship for
Yang-Mills fields.

To explore such a possibility, we note first that in the alternative
formulation of pure electrodynamics as summarised in the second column of
Chart I, the `dual potential' emerged spontaneously as the Lagrange multiplier
for a constraint imposed on the variable $F_{\mu\nu}(x)$ to guarantee the
existence of the ordinary potential $A_\mu(x)$.  This suggests then that
perhaps in nonabelian Yang-Mills theories, a generalised dual potential will
also emerge as the Lagrange multiplier to a similar constraint.  A possible
strategy would thus be to examine how this alternative formulation of pure
electrodynamics can be extended to nonabelian Yang-Mills theories.

With this view in mind, one notices immediately some important differences
between the abelian and nonabelian theories.  First, $F_{\mu\nu}(x)$ is gauge
invariant in electromagnetism but only covariant in nonabelian Yang-Mills
theory.  Second, $F_{\mu\nu}(x)$ fully describes the classical abelian theory,
but for the nonabelian theory, it is known that several gauge inequivalent
potentials can correspond to the same $F_{\mu\nu}(x)$, meaning that
$F_{\mu\nu}(x)$ is insufficient to describe fully even the classical Yang-Mills
field.  Third, the condition (\ref{gausslaw}) guarantees in the abelian theory
that $F_{\mu\nu}(x)$ is a gauge field derivable from a potential, but for
the nonabelian theory, the corresponding condition:
\begin{equation}
D_\nu \mbox{\mbox{}$^*\!$} F^{\mu\nu}(x) = 0,
\label{xGausslaw}
\end{equation}
cannot even be given a meaning until the potential $A_\mu(x)$ is defined.
Thus, in order to mimic the above `alternative formulation' of
electromagnetism,
one has first to find some suitable gauge invariant set of field variables
which can fully describe the theory, and then to impose on them appropriate
constraints so as to enable one to recover the gauge potentials.

A solution to this problem has been found using loop variables which we
summarise as follows.  We begin by defining {\it parametrised loops} as maps
of the circle $S^1$ to 4-dimensional space-time, thus:
\begin{equation}
\xi: \{\xi^\mu(s); s = 0 \rightarrow 2\pi, \xi^\mu(0) = \xi^\mu(2\pi)
   = \xi^\mu_0 \}.
\label{xiofs}
\end{equation}
We need to consider only those loops beginning and ending at some fixed
reference point $P_0 = \{\xi^\mu_0\}$.  For each parametrised loop $\xi$,
one can then define a phase factor (Dirac factor or Wilson loop):
\begin{equation}
\Phi[\xi] = P_s \exp ig \int_0^{2\pi} ds A_\mu(\xi(s)) \dot{\xi}^\mu(s),
\label{Phiofxi}
\end{equation}
where $P_s$ denotes ordering, say from right to left, and a dot differentiation
in the loop parameter $s$.  Following Polyakov \cite{polyakov}, define next:
\begin{equation}
F_\mu[\xi|s] = \frac{i}{g} \Phi^{-1}[\xi] \delta_\mu(s) \Phi[\xi],
\label{Fmuxis}
\end{equation}
where we denote by $\delta_\mu(s)$ the ordinary functional derivative
$\delta/
\delta \xi^\mu(s)$ with respect to $\xi^\mu(s)$.\footnote{Explicitly, for
any functional $\Psi[\xi]$ of $\xi$, the functional derivative is defined as:
\begin{equation}
\frac{\delta}{\delta \xi^\mu(s)} \Psi[\xi] = \lim_{\Delta \rightarrow 0}
   \frac{1}{\Delta} \{\Psi[\xi'] - \Psi[\xi]\},  \nonumber
\end{equation}
with:
\begin{equation}
\xi'^\alpha(s') = \xi^\alpha(s') + \Delta \delta^\alpha_\mu \delta(s-s')
   \nonumber,
\end{equation}
where in case of ambiguity, the variation is to be evaluated by first
replacing $\delta(s-s')$ with a smooth function of $s'$ peaked at $s$ with
width $2 \eta$, and then taking the $\eta \rightarrow 0$ limit afterwards.}
This quantity $F_\mu[\xi|s]$, which may be thought of as a `connection' giving
parallel phase transport in loop space, is what is proposed as the field
variable for Yang-Mills fields analogous to $F_{\mu\nu}(x)$ above for
electromagnetism.

In terms of ordinary local field variables, $F_\mu[\xi|s]$ can be expressed as:
\begin{equation}
F_\mu[\xi|s] = \Phi_\xi^{-1}(s,0) F_{\mu\nu}(\xi(s)) \Phi_\xi(s,0)
   \dot{\xi}^\nu(s),
\label{Fmuxisinx}
\end{equation}
where:
\begin{equation}
\Phi_\xi(s_2,s_1) = P_s \exp ig \int_{s_1}^{s_2} ds A_\mu(\xi(s))
   \dot{\xi}^\mu(s),
\label{Phixis2s1}
\end{equation}
is the parallel phase transport from $s_1$ to $s_2$.  From (\ref{Fmuxisinx})
one sees that $F_\mu[\xi|s]$ is closely related
to the local field tensor $F_{\mu\nu}(x)$, and indeed when the theory is
abelian, it reduces just to the field tensor dotted into the tangent to the
loop $\xi$ at the point labelled by the parameter $s$.  Further, it is obvious
that $F_\mu[\xi|s]$ is gauge invariant apart from an $x$-independent rotation
at $P_0$, which can be easily handled and will henceforth be ignored.  And
that $F_\mu[\xi|s]$ is sufficient for a full description of the Yang-Mills
theory will be clear when one shows that the gauge potential $A_\mu(x)$ is
recoverable from $F_\mu[\xi|s]$.  One sees thus that $F_\mu[\xi|s]$ does
have the credentials we sought for the field variable in the gauge invariant
`alternative' formulation of the Yang-Mills theory.

However, there are, of course, vastly more loops in space-time than there
are points, so that the variables $\{F_\mu[\xi|s]\}$ which are labelled
by loops $\xi$ as well as points $s$ on the loops must be highly redundant
when, as one recalls, the theory is already sufficiently described by
$\{A_\mu(x)\}$ which are labelled only by the points $x$ in space-time.
To use $\{F_\mu[\xi|s]\}$ as variables, they must therefore be
severely constrained so as to remove their redundancy, i.e. to ensure that
they are all in fact derivable from a potential $A_\mu(x)$ as per
(\ref{Fmuxis})
and (\ref{Phiofxi}).  In other word, we need a parallel to (\ref{gausslaw})
of the abelian theory which removed the redundancy in $\{F_{\mu\nu}(x)\}$
and ensured that they can be derived from a potential.  This problem has also
been solved by a result which we shall refer to here as the {\it Extended
Poincar\'e Lemma}. \cite{annphys} This affirms that an $\{A_\mu(x)\}$ can
be recovered from a given set $\{F_\mu[\xi|s]\}$ provided that the latter
satisfies the following condition:
\begin{equation}
G_{\mu\nu}[\xi|s, s'] = 0,
\label{Gausslaw}
\end{equation}
where:
\begin{equation}
G_{\mu\nu}[\xi|s,s'] = \delta_\nu(s') F_\mu[\xi|s] - \delta_\mu(s)
F_\nu[\xi|s']
   + ig [F_\mu[\xi|s], F_\nu[\xi|s']];
\label{Gmunu}
\end{equation}
and provided that $F_\mu[\xi|s]$ is transverse, namely:
\begin{equation}
F_\mu[\xi|s] \dot{\xi}^\mu(s) = 0,
\label{Ftransverse}
\end{equation}
and does not depend on $\xi(s')$ for $s' > s$.  Conversely, it is readily
seen that if $F_\mu[\xi|s]$ is derivable from a potential $A_\mu(x)$ in the
manner (\ref{Fmuxis}) and (\ref{Phiofxi}), then the above three conditions
are automatically satisfied.

{}From (\ref{Fmuxis}), one sees that the condition (\ref{Ftransverse})
just says that the longitudinal derivative of $\Phi[\xi]$ with respect to
$\xi(s)$ is zero, or that $\Phi[\xi]$ is independent of the parametrisation
of $\xi$.  Further, from (\ref{Fmuxisinx}), (\ref{Ftransverse}) is seen to
be equivalent just to the statement in the abelian theory that $F_{\mu\nu}(x)$
is antisymmetric under an interchange of its indices.  On the other hand,
the condition that $F_\mu[\xi|s]$ does not depend on $\xi(s')$ for $s' > s$
is a consequence of the ordering $P_s$ of the exponential in (\ref{Phiofxi}),
and is related to the fact in the abelian theory that $F_{\mu\nu}(x)$
is a local quantity depending only on $x$.  Both these conditions being
relatively easy to handle, they will henceforth be regarded as understood
and absorbed into the notation for $F_\mu[\xi|s]$.

The other condition (\ref{Gausslaw}) is more interesting, deserving some
scrutiny.  Recalling that $F_\mu[\xi|s]$ can be regarded as a `connection'
for parallel phase transport in loop space, one sees that
$G_{\mu\nu}[\xi|s,s']$
is just the corresponding curvature, which as usual represents the change
in phase obtained in parallelly transporting around an infinitesimal closed
circuit.  In ordinary space, this circuit in loop space may be pictured as
a surface where for $s \neq s'$, $G_{\mu\nu}[\xi|s,s']$ is seen to be zero
since the surface encloses no volume, but for $s' = s$, the surface
becomes a sort of an infinitesimal version of that $\Sigma$ which
was used to specify a monopole of the abelian theory in the preceding section.
Indeed, the same procedure has been used by Lubkin \cite{lubkin},
Wu and Yang \cite{patching} and Coleman \cite{coleman} to specify a monopole
of  Yang-Mills theories in general, as will be discussed in further detail
later in Section 5.  The condition ({\ref{Gausslaw}) is thus seen to be just
the statement that there are no monopoles anywhere in the nonabelian theory,
and has thus exactly the same physical content as the Bianchi identity
(\ref{gausslaw}) of electromagnetism.  This has enabled us to apply the
Poincar\'e lemma in the abelian case to remove the redundancy of the
variables $F_{\mu\nu}(x)$ and at the same time to recover the potential
$A_\mu(x)$, just as (\ref{Gausslaw}) is meant to do here for the
variables $F_\mu[\xi|s]$.  One sees therefore that there is indeed a close
parallel between the so-called Extended Poincar\'e Lemma above and the
original Poincar\'e lemma in the abelian theory which justifies the above
choice of nomenclature.

With this result, we can now proceed to reformulate Yang-Mills theory in
terms of $F_\mu[\xi|s]$.  Using (\ref{Fmuxisinx}), the action ${\cal A}^0_F$
in (\ref{ScriptA0F}) can be rewritten as:
\begin{equation}
{\cal A}^0_F = - \frac{1}{4 \pi {\bar N}} \int \delta \xi ds \,\mbox{\rm Tr}
   \{F_\mu[\xi|s] F^\mu[\xi|s]\} \dot{\xi}(s)^{-2},
\label{ScriptA0FL}
\end{equation}
where the integral is to be taken over all parametrised loops \footnote{We
note that parametrised loops $\xi$ being by definition just functions of
$s$, integrals over $\xi$ are just ordinary functional integrals, which is
in fact one reason why we prefer to work with parametrised loops rather
than the actual loops.} $\xi$ and over all points $s$ on each loop, and
${\bar N}$ is an (infinite) normalisation factor defined as:
\begin{equation}
{\bar N} = \int_0^{2\pi} ds \int \prod_{s' \neq s} d^4\xi(s').
\label{Nbar}
\end{equation}
Equations of motion are to be obtained by extremising (\ref{ScriptA0FL})
with respect to $F_\mu[\xi|s]$ under the constraint (\ref{Gausslaw}).  Using
as Lagrange multipliers $L_{\mu\nu}[\xi|s]$, we extremise:
\begin{equation}
{\cal A}_F = {\cal A}^0_F + \int \delta \xi ds \,\mbox{\rm Tr}
\{L_{\mu\nu}[\xi|s]
   G^{\mu\nu}[\xi|s]\},
\label{ScriptAFL}
\end{equation}
with respect to $F_\mu[\xi|s]$. Notice that $G_{\mu\nu}[\xi|s,s']$ being
automatically zero for $s \neq s'$, we need to impose the constraint only
for $s = s'$, and the symbol $G_{\mu\nu}[\xi|s]$ in (\ref{ScriptAFL}) is
meant to represent $G_{\mu\nu}[\xi|s,s']$ with $s'$ smeared around $s$.

The extremisation gives:
\begin{equation}
F_\mu[\xi|s] = -(4\pi{\bar N}\dot{\xi}(s)^2) {\cal D}^\nu(s) L_{\mu\nu}[\xi|s],
\label{FmuxisinL}
\end{equation}
where ${\cal D}_\mu(s)$ is the `covariant derivative' in loop space:
\begin{equation}
{\cal D}_\mu(s) = \delta_\mu(s) -ig [F_\mu[\xi|s], \ \ \ ].
\label{covderivL}
\end{equation}
(\ref{FmuxisinL}) is the equation of motion in parametric form, from which
the Lagrange multipliers $L_{\mu\nu}[\xi|s]$ can be eliminated \cite{physrev}
to give:
\begin{equation}
{\cal D}^\mu(s) F_\mu[\xi|s] = 0.
\label{prepolyakov}
\end{equation}
Further, by (\ref{covderivL}), one sees that (\ref{prepolyakov}) is the same
as:
\begin{equation}
\delta^\mu(s) F_\mu[\xi|s] = 0,
\label{polyakov}
\end{equation}
the Polyakov equation, which is known, by (\ref{Fmuxisinx}), to be exactly
equivalent to the Yang-Mills equation (\ref{yangmillseq}).  The reformulation
of the original Yang-Mills problem in terms of the loop variables
$F_\mu[\xi|s]$
is now complete.

We next recall that the original reason given above for attempting a
reformulation of the theory in terms of $F_\mu[\xi|s]$ was the suspicion that,
by analogy with the parallel formulation of the abelian theory as summarised
in the second column of Chart I, the Lagrange multiplier $L_{\mu\nu}[\xi|s]$
should play here the role of a dual potential, given a suitable generalisation
of the concept of duality.  That this is indeed possible can be made more
transparent by a few notational rearrangements.  We note first that     .
(\ref{FmuxisinL}) can be rewritten as:
\begin{equation}
F_\mu[\xi|s] = -(4\pi{\bar N}\dot{\xi}(s)^2) \Phi^{-1} \delta^\nu(s)
   \{\Phi L_{\mu\nu}[\xi|s] \Phi^{-1}\} \Phi,
\label{FmuxisinLp}
\end{equation}
for any $\Phi$ satisfying:
\begin{equation}
\delta_\mu(s) \Phi = - ig \Phi F_\mu[\xi|s],
\label{diffPhi}
\end{equation}
which holds for any $\Phi_\xi(s_1, 0)$ defined as in (\ref{Phixis2s1}) with
$s_1 > s$.  In particular, we may choose $\Phi$ to be $\Phi_\xi(s_+, 0)$,
where $s_+$ is $s + \eta$ with $2 \eta > 0$ being the `width' of the
$\delta$-function in the definition of the functional derivative
$\delta^\nu(s)$, which `width' is to be taken to zero afterwards, as explained
in the footnote after (\ref{Fmuxis}). Hence, if one defines:
\begin{equation}
H_{\mu\nu\rho}[\xi|s] = - \frac{1}{\sqrt{6}}\epsilon_{\mu\nu\rho\sigma}
   \Phi_\xi(s, 0) F^\sigma[\xi|s] \Phi_\xi^{-1}(s, 0),
\label{Hmunurho}
\end{equation}
as the (generalised) dual to $F_\mu[\xi|s]$, then by (\ref{FmuxisinLp})
and (\ref{diffPhi}) it is seen to be expressible as:
\begin{equation}
H_{\mu\nu\rho}[\xi|s] = \delta_\mu(s) T_{\nu\rho}[\xi|s] + \delta_\nu(s)
   T_{\rho\mu}[\xi|s] + \delta_\rho(s) T_{\mu\nu}[\xi|s],
\label{HmunurhoinT}
\end{equation}
for:
\begin{equation}
T_{\mu\nu}[\xi|s] = - \sqrt{\frac{2}{3}} \pi \bar{N} \dot{\xi}(s)^2
   \epsilon_{\mu\nu\rho\sigma} \Phi_\xi(s_+, 0) L^{\rho\sigma}[\xi|s]
   \Phi_\xi^{-1}(s_+, 0).
\label{Tmunu}
\end{equation}
In other word, this dual $H$ to $F$ is here expressed as an exterior derivative
of $T$, which we can thus regard as a generalised dual potential.

For the abelian theory, by (\ref{Fmuxisinx}),
(\ref{Hmunurho}) reduces to
\begin{equation}
H_{\mu\nu\rho}[\xi|s] = - \frac{1}{\sqrt{6}}\epsilon_{\mu\nu\rho\sigma}
   F^{\sigma\alpha} (\xi(s)) \dot{\xi}_\alpha(s),
\label{Hmunurhoab}
\end{equation}
so that:
\begin{equation}
\mbox{\mbox{}$^*\!$} F_{\mu\nu}(x) = - \frac{2 \sqrt{6}}{\bar{N}} \int \delta
\xi ds
   H_{\mu\nu\rho}[\xi|s] \dot{\xi}^\rho(s) \dot{\xi}(s)^{-2} \delta(x-\xi(s)),
\label{fstarinH}
\end{equation}
while (\ref{HmunurhoinT}) reduces to (\ref{dfmunu}) with:
\begin{equation}
{\tilde A}_\mu(x) = \frac{2 \sqrt{6}}{\bar{N}} \int \delta \xi ds
   T_{\mu\nu}[\xi|s] \dot{\xi}^\nu(s) \dot{\xi}(s)^{-2} \delta(x-\xi(s)).
\label{atinT}
\end{equation}
Hence, apart from the fact that in the loop space representation quantities
such as $H_{\mu\nu\rho}[\xi|s], T_{\mu\nu}[\xi|s]$ and $F_\mu[\xi|s]$ may
carry one more or one less index than their corresponding local quantities
$\mbox{\mbox{}$^*\!$} F_{\mu\nu}(x), A_\mu(x)$ and $F_{\mu\nu}(x)$ due to the
extra
direction carried by the tangent to the loop, the equations (\ref{Hmunurho})
and (\ref{HmunurhoinT}) are just the generalisations of respectively the
abelian duality relationship (\ref{fstar}) and the definition (\ref{dfmunu})
of the dual potential.

Furthermore, the condition (\ref{HmunurhoinT}) implies that:
\begin{equation}
\delta^\sigma \epsilon_{\mu\nu\rho\sigma} H^{\mu\nu\rho}[\xi|s] = 0,
\label{divstarH}
\end{equation}
as can be shown by direct calculation, and by (\ref{Hmunurho}), this is seen
to be identical to the Polyakov equation (\ref{polyakov}).  Hence, the
existence of the dual potential $T_{\mu\nu}[\xi|s]$ guarantees the validity
of the equation of motion, which is the parallel to the statement that
(\ref{dfmunu}) guarantees (\ref{maxwelleq}) in the abelian theory.  Conversely,
since the left-hand side of (\ref{divstarH}) can be regarded as the exterior
derivative of the 3-form $H$, its vanishing implies via the Poincar\'e lemma
that there exists a 2-form $T$ of which $H$ is itself the exterior derivative.
In other words, the equation of motion (\ref{polyakov}) or (\ref{divstarH})
also implies (\ref{HmunurhoinT}), which is the parallel to the statement that
(\ref{maxwelleq}) implies (\ref{dfmunu}) in the abelian case.

For electromagnetism, the tensor $\mbox{\mbox{}$^*\!$} F_{\mu\nu}(x)$ as given
in
(\ref{dfmunu}) was invariant under the transformation (\ref{uttransf}).
Similarly here, the tensor $H_{\mu\nu\rho}[\xi|s]$ as given by
(\ref{HmunurhoinT}) is invariant under the transformation:
\begin{equation}
T_{\mu\nu}[\xi|s] \longrightarrow T_{\mu\nu}[\xi|s] + \delta_{[\nu}
   {\tilde \Lambda}_{\mu]}.
\label{Uttransf}
\end{equation}
Moreover, under the transformation (\ref{uttransf}), the abelian action
(\ref{scriptAF}) is invariant, as can be seen after an integration by parts
from the antisymmetry of $\mbox{\mbox{}$^*\!$} F_{\mu\nu}(x)$.  Under the
analogous transformation (\ref{Uttransf}) here, the Lagrange multiplier
$L_{\mu\nu}[\xi|s]$ transforms as:
\begin{equation}
L_{\mu\nu}[\xi|s] \longrightarrow L_{\mu\nu}[\xi|s] + \sqrt{6} (4\pi \bar{N}
   \dot{\xi}(s)^2)^{-1} \epsilon_{\mu\nu\rho\sigma} {\cal D}^\sigma(s)
   \{\Phi_\xi^{-1}(s_+,0) \tilde{\Lambda}^\rho[\xi|s] \Phi_\xi(s_+,0) \}.
\label{UttransfL}
\end{equation}
We note that, apart from some trivial factors and changes in notation,
this is exactly the transformation of the antisymmetric tensor potential
suggested by Freedman and Townsend \cite{freedtown} in a different context, and
that the action (\ref{ScriptAFL}) has also the form of their invariant action.
Indeed, on substituting into (\ref{ScriptAFL}) and integrating by parts with
respect to $\delta \xi(s)$, this gives an increment:
\begin{equation}
\Delta {\cal A}_F = \sqrt{6} (4\pi \bar{N} \dot{\xi}(s)^2)^{-1}
   \int \delta \xi ds \,\mbox{\rm Tr}\{\Phi_\xi^{-1}(s_+,0) {\tilde
\Lambda}_\sigma[\xi|s]
  \Phi_\xi(s_+,0)\epsilon^{\mu\nu\rho\sigma}{\cal
D}_\rho(s)G_{\mu\nu}[\xi|s]\},
\label{DeltaAF}
\end{equation}
which vanishes because of the Bianchi identity satisfied by $G_{\mu\nu}[\xi|s]$
by virtue of its definition in (\ref{Gmunu}).  Hence, we conclude that,
analogously to the abelian theory, the symmetry in the nonabelian theory
is also enlarged by a new symmetry on the dual potential.  That this new
symmetry should form another $SU(N)$, however, will only be apparent later
when considering the interaction of the Yang-Mills field with a colour
magnetic charge.

For easier comparison with electromagnetism, the arguments in this section
are summarised on the left half of Chart III.  As can be seen, the analogy
with Chart I for the abelian theory is rather close.

\setcounter{equation}{0}
\section{Pure Yang-Mills Theory - Dual Formulation}

In electromagnetism, dual symmetry means that electricity and magnetism can
be inverted so that one can reformulate the theory entirely in terms of
$\mbox{\mbox{}$^*\!$} F_{\mu\nu}(x)$ and ${\tilde A}_\mu(x)$ as variables
rather than
$F_{\mu\nu}(x)$ and $A_\mu(x)$.  Hence, in the nonabelian theory, having
introduced now a generalised dual relationship with $H_{\mu\nu\rho}[\xi|s]$
and $T_{\mu\nu}[\xi|s]$ as respectively the `dual field' and `dual potential',
one would hope to be able also to reformulate the theory in a dual fashion,
starting with $H_{\mu\nu\rho}[\xi|s]$ and $T_{\mu\nu}[\xi|s]$ as variables
instead of $F_\mu[\xi|s]$ and $A_\mu(x)$.  Further, in analogy to the direct
formulation, one would expect that the formulation can be carried out in two
ways: either by expressing the action in terms of the potential
$T_{\mu\nu}[\xi|s]$ and extremising with respect to $T_{\mu\nu}[\xi|s]$, or
else by adopting the ${\tilde U}$-invariants $H_{\mu\nu\rho}[\xi|s]$ as
variables but imposing appropriate constraints on them so as to ensure that
they can be derived from a potential.  In either case, one would expect to
obtain exactly the same result as in the direct formulation of the theory.

In terms of the variables $H_{\mu\nu\rho}[\xi|s]$, the field action
(\ref{ScriptA0F}) or (\ref{ScriptA0FL}) can be rewritten as:
\begin{equation}
{\cal A}^0_F = \frac{1}{4 \pi \bar{N}} \int \delta \xi ds \,\mbox{\rm Tr}
   \{ H_{\mu\nu\rho}[\xi|s] H^{\mu\nu\rho}[\xi|s] \} \dot{\xi}(s)^{-2}.
\label{ScriptAt0FL}
\end{equation}
Adopting the first approach, we substitute the expression (\ref{HmunurhoinT})
into (\ref{ScriptAt0FL}) and extremise with respect to $T_{\mu\nu}[\xi|s]$.
This gives:
\begin{equation}
\delta_\rho(s) H^{\mu\nu\rho}[\xi|s] = 0,
\label{GausslawH}
\end{equation}
as the equation of motion.  In analogy to the direct formulation of pure
Yang-Mills theory as summarised in the first column of Chart III, as well
as to the dual formulation of the abelian theory as summarised in the last
column of Chart I, one would expect this equation (\ref{GausslawH}) to
guarantee the existence of the (direct) potential $A_\mu(x)$.  That this
is the case can be seen as follows.  The loop space curvature
$G_{\mu\nu}[\xi|s,s']$ in (\ref{Gmunu}) can be expressed as:
\begin{eqnarray}
G_{\mu\nu}[\xi|s,s']& = &\Phi_\xi^{-1}(s,0) \delta_\nu(s') \{ \Phi_\xi(s,0)
   F_\mu[\xi|s] \Phi_\xi^{-1}(s,0) \} \Phi_\xi(s,0) \nonumber \\
   & & - \Phi_\xi^{-1}(s',0) \delta_\mu(s) \{ \Phi_\xi(s',0) F_\nu[\xi|s']
   \Phi_\xi^{-1}(s',0) \} \Phi_\xi(s',0),
\label{Gmununewex}
\end{eqnarray}
since, by (\ref{diffPhi}):
\begin{eqnarray}
& \delta_\nu(s') \{\Phi_\xi(s,0) F_\mu[\xi|s] \Phi_\xi^{-1}(s,0) \}
   \nonumber \\  &
   =\Phi_\xi(s,0) \{\delta_\nu(s') F_\mu[\xi|s]-ig[F_\nu[\xi|s'],F_\mu[\xi|s]]
   \theta(s-s')\} \Phi_\xi^{-1}(s,0),
\label{delFmuhat}
\end{eqnarray}
and a similar relation with $s \leftrightarrow s', \mu \leftrightarrow \nu$.
Hence, by (\ref{Hmunurho}), we have, on putting $s = s'$:
\begin{equation}
G_{\mu\nu}[\xi|s] = - \sqrt{\frac{3}{2}} \epsilon_{\mu\nu\rho\sigma}
   \Phi_\xi^{-1}(s,0) \delta_\alpha(s) H^{\rho\sigma\alpha}[\xi|s]
   \Phi_\xi(s,0) = 0,
\label{GausslawHp}
\end{equation}
which implies by the Extended Poincar\'e Lemma in (\ref{Gausslaw}) that
$A_\mu(x)$ exists and is related to $H_{\mu\nu\rho}[\xi|s]$ by
(\ref{Hmunurho}),
(\ref{Fmuxisinx}) and (\ref{Phixis2s1}) as required.  We have thus completed
the analogy here to the direct formulation summarised in the first column of
Chart III.

Suppose we wish now to adopt instead the ${\tilde U}$-invariants
$H_{\mu\nu\rho}[\xi|s]$ as variables, which, as we have seen in the abelian
theory, are more convenient when we later deal with monopoles.  In that case,
we need to impose on $H_{\mu\nu\rho}[\xi|s]$ appropriate constraints so as
to ensure that $H_{\mu\nu\rho}[\xi|s]$ is indeed derivable from a potential
$T_{\mu\nu}[\xi|s]$ as it should.  The appropriate constraints are just those
exhibited in (\ref{divstarH}) because of the Poincar\'e lemma in loop space,
as was explained there.  This is the dual image of the
constraint $G_{\mu\nu}[\xi|s] = 0$ in the direct formulation (Chart III, second
column) and the parallel of the constraint $\partial_\nu F^{\mu\nu}(x) = 0$
in the dual formulation of the abelian theory (Chart I, third column).

Hence, we write:
\begin{equation}
{\tilde {\cal A}}_F = {\tilde {\cal A}}_F^0 + \int \delta \xi ds \,\mbox{\rm
Tr}
  \{ L[\xi|s] \delta_\mu \epsilon^{\mu\nu\rho\sigma}
H_{\nu\rho\sigma}[\xi|s]\},
\label{ScriptAtFL}
\end{equation}
with ${\tilde {\cal A}}_F^0$ given as in (\ref{ScriptAt0FL}), and $L[\xi|s]$
as the Lagrange multiplier.  This is to be extremised with respect to
$H_{\mu\nu\rho}[\xi|s]$, which gives as the equation of motion:
\begin{equation}
H_{\mu\nu\rho}[\xi|s] = -(2\pi \bar{N} \dot{\xi}(s)^2)
   \epsilon_{\mu\nu\rho\sigma} \delta^\sigma L[\xi|s],
\label{HmunurhoinL}
\end{equation}
in terms of the Lagrange multiplier $L[\xi|s]$ as parameter.  This is the
dual image of (\ref{FmuxisinL}) in the direct formulation.

Extending the analogy further, one would expect that (\ref{HmunurhoinL})
would imply the equation of motion (\ref{GausslawH}) derived before, and that
the Lagrange multiplier $L[\xi|s]$ would play the role of the `dual potential'
for $H_{\mu\nu\rho}[\xi|s]$, or in other words the potential for
$F_\mu[\xi|s]$,
namely $A_\mu(x)$.  That (\ref{HmunurhoinL}) implies (\ref{GausslawH}) is
straightforward, since:
\begin{equation}
\delta_\rho H^{\mu\nu\rho}[\xi|s] = -(2\pi \bar{N} \dot{\xi}(s)^2)
   \epsilon^{\mu\nu\rho\sigma} \delta_\rho(s) \delta_\sigma(s) L[\xi|s],
\label{divHinL}
\end{equation}
vanishes by symmetry, and as already noted,
this implies via the vanishing of $G_{\mu\nu}[\xi|s]$ that a potential
$A_\mu(x)$ exists.  That this $A_\mu(x)$ should be given in terms of the
Lagrange multiplier $L[\xi|s]$ is a little more involved but can be seen
as follows.

Let us first, for convenience, normalise $L[\xi|s]$ by a factor,
calling it:
\begin{equation}
W[\xi|s] = \sqrt{6} (2\pi \bar{N} \dot{\xi}(s)^2) L[\xi|s],
\label{WinL}
\end{equation}
so that by (\ref{divHinL}) above:
\begin{equation}
F_\mu[\xi|s] = \Phi_\xi^{-1}(s,0) \delta_\mu(s) W[\xi|s] \Phi_\xi(s,0).
\label{FmuxisinW}
\end{equation}
Define then:
\begin{equation}
A_\mu(x) = \frac{2}{\bar{N}} \int \delta \xi ds W[\xi|s] \dot{\xi}_\mu(s)
   \dot{\xi}(s)^{-2} \delta(x-\xi(s)),
\label{AmuinW}
\end{equation}
which we wish to show is the actual potential we want.  With this $A_\mu(x)$
in (\ref{AmuinW}) as connection, one can define as usual by (\ref{Phiofxi})
a phase factor $\Phi[\xi]$ for any parametrised closed loop $\xi$ beginning
and ending at the reference point $P_0$.  For every point $x$ in space-time,
choose a path $\gamma_x$ linking $x$ to $P_0$ in such a way that neighbouring
points will be linked to $P_0$ by neighbouring paths.  Thus, in particular,
$\gamma_x$ may be chosen just as the straight line joining $x$ to $P_0$.
Given any loop $\xi$ then, the two points on it, $\xi(s_+)$ and $\xi(s_-)$
for $s_\pm = s \pm \eta$, specify a wedge-shaped closed loop
$\gamma_{\xi(s_+)}^{-1} \circ \xi(s_+,s_-) \circ \gamma_{\xi(s_-)}$, as
depicted in Figure \ref{Wedge}, where $\xi(s_+,s_-)$ denotes the segment
of $\xi$ between $s_-$ and $s_+$.  For this closed loop, one can construct
\begin{figure}
\vspace{8cm}
\caption{Illustration for the recovery of $A_\mu(x)$ from $W[\xi|s]$}
\label{Wedge}
\end{figure}
the standard phase factor (\ref{Phiofxi}), an element of the gauge group
which we denote as:
\begin{equation}
g(\xi(s_+),\xi(s_-)) = \Phi[\gamma_{\xi(s_+)}^{-1} \circ \xi(s_+,s_-)
   \circ \gamma_{\xi(s_-)} ].
\label{gxis+xis-}
\end{equation}
Then, by the standard loop space procedure for constructing local potentials,
we conclude that:
\begin{equation}
A_\mu(\xi(s)) \dot{\xi}^\mu(s) \sim -\frac{i}{g} \lim_{\eta \rightarrow 0}
   \frac{1}{2 \eta} \{ g(\xi(s_+), \xi(s_-)) - 1\}
\label{Amuing}
\end{equation}
where $\sim$ means that the two sides are equal up to an ordinary Yang-Mills
gauge transformation.  [An outline of a proof for (\ref{Amuing}) can be found,
for example, in ref. \cite{annphys}]  Further, as is well known, a gauge
transformation
here is the same as a choice of the path $\gamma_x$ linking each point $x$
in space-time to the reference point $P_0$.  Hence by the appropriate choice
of $\gamma_x$, which we know exist but need not specify, we can actually
replace the sign $\sim$ above by an equality and write instead:
\begin{equation}
A_\mu(\xi(s)) \dot{\xi}^\mu(s) = -\frac{i}{g} \lim_{\eta \rightarrow 0}
   \frac{1}{2 \eta} \{g(\xi(s_+), \xi(s_-)) - 1\},
\label{Amuinga}
\end{equation}
where, we recall, $g(\xi(s_+), \xi(s_-))$ is the phase factor $\Phi$ in
(\ref{Phiofxi}) for the wedge-shaped loop $\gamma_{\xi(s_+)}^{-1} \circ
\xi(s_+, s_-) \circ \gamma_{\xi(s_-)}$.  If we put then:
\begin{equation}
W[\xi|s] = -\frac{i}{g} \frac{1}{\eta} \{ g(\xi(s_+), \xi(s_-)) - 1 \},
\label{Wing}
\end{equation}
we see we recover $A_\mu(x)$ as defined in (\ref{AmuinW}).

We have yet to show that the $W[\xi|s]$ so constructed does indeed satisfy
the equation (\ref{FmuxisinW}).
This is best seen diagrammatically.  Differentiating the wedge in Figure
\ref{Wedge} with respect to $\xi^\mu(s)$, one obtains Figure \ref{diffwedge}.
\begin{figure}
\vspace{8cm}
\caption{Differentiating the wedge}
\label{diffwedge}
\end{figure}
Multiplying next by the factors $\Phi_\xi^{-1}(s_+,0)$ and $\Phi_\xi(s_+,0)$
gives Figure \ref{Ffromwedge}, which is exactly $F_\mu[\xi|s]$ as defined in
\begin{figure}
\vspace{8cm}
\caption{Obtaining $F_\mu[\xi|s]$ by differentiating the wedge}
\label{Ffromwedge}
\end{figure}
(\ref{Fmuxis}).  This completes then our demonstration that the quantity
defined in (\ref{AmuinW}) in terms of $W[\xi|s]$ or $L[\xi|s]$ is indeed
the potential $A_\mu(x)$ we sought.

We have shown now not only that the action when constrained by (\ref{divstarH})
leads to the correct equation of motion but also that the Lagrange multiplier
$L[\xi|s]$ gives via (\ref{AmuinW}) the local potential, in parallel to
(\ref{ainlambdat}) for the abelian theory.  It remains to show only that the
action ${\tilde {\cal A}}_F$ is invariant under a gauge transformation of
$A_\mu(x)$.  As usual, of course, $A_\mu(x)$ transforms under an infinitesimal
gauge transformation as:
\begin{equation}
A_\mu(x) \longrightarrow A_\mu(x) + \partial_\mu \Lambda(x) + ig[\Lambda(x),
   A_\mu(x)].
\label{UtransfAmu}
\end{equation}
Under such a transformation, $F_\mu[\xi|s]$ is invariant, and $\Phi_\xi(s,0)$
transforms as:
\begin{equation}
\Phi_\xi(s,0) \longrightarrow [1 + ig \Lambda(\xi(s))] \Phi_\xi(s,0)
\label{UtransfPhis0}
\end{equation}
so that $H_{\mu\nu\rho}[\xi|s]$ is covariant, which implies that
${\tilde {\cal A}}_F^0$ in (\ref{ScriptAt0FL}) is invariant under this
$U$-transformation.  For ${\tilde {\cal A}}_F$ in
(\ref{ScriptAtFL}) to be invariant, we want therefore that the second term
in (\ref{ScriptAtFL}) due to the constraint be also invariant.  From
(\ref{Wing}), and recalling the general transformation properties of phase
factors, one sees that $W[\xi|s]$ transforms as:
\begin{equation}
W[\xi|s] \longrightarrow \{1 + ig \Lambda(\xi(s_+))\} W[\xi|s]
   \{1 - ig \Lambda(\xi(s_-))\} + \frac{1}{\eta} \{\Lambda(\xi(s_+))
   - \Lambda(\xi(s_-))\}.
\label{UtransfW}
\end{equation}
Substituting this into the second term of ${\tilde {\cal A}}_F$ in
(\ref{ScriptAtFL}) and performing an integration by parts with respect to
$\xi$, one sees that the derivative $\delta_\sigma(s)$ does not operate
on the local quantities $\Lambda(\xi(s_\pm))$ defined at the end-points of
the interval $(s_+, s_-)$ so that the second term in (\ref{UtransfW}) gives
zero variation.  One obtains then just:
\begin{eqnarray}
- \int \delta \xi ds&\,\mbox{\rm Tr}&\{[1+ig \Lambda(\xi(s_+))]
\delta_\sigma(s) L[\xi|s]
   [1-ig \Lambda(\xi(s))] \nonumber \\
   & \times & [1+ig \Lambda(\xi(s))] \epsilon_{\mu\nu\rho\sigma}
   H^{\mu\nu\rho}[\xi|s] [1-ig \Lambda(\xi(s))] \}
\label{UtransfAItFL}
\end{eqnarray}
which gives the same value as before when $s_+ \rightarrow s_-$.  The action
${\tilde {\cal A}}_F$ is thus indeed invariant as required.

The result of this section is summarised on the right-hand side of Chart III,
which is seen to be quite a close dual image of the direct formulation on the
left, and closely parallels also the dual formulation of the abelian theory
on the right of Chart I.

\setcounter{equation}{0}
\section{Yang-Mills Field with Colour Magnetic Charge}

By a {\it colour magnetic charge} we mean here just a monopole of the
Yang-Mills
field in the same way that a magnetic charge in electromagnetism is understood
as a monopole of the Maxwell field, but since the term `monopole' has been used
to mean quite different things in a wide variety of contexts, we need to be
more specific.  Let us first recall the Lubkin \cite{lubkin},
Wu-Yang \cite{patching}, Coleman \cite{coleman} definition of a
monopole as a topological obstruction in a Yang-Mills field.  As for the
abelian theory, start again with
a family of closed loops enveloping a surface $\Sigma$ as that considered in
Section 2, and for each loop $C_t$, suitably parametrised by a function,
say $\xi_t^\mu(s): s = 0 \rightarrow 2\pi$, construct the phase factor
$\Phi[\xi_t]$ as in (\ref{Phiofxi}).  For each $t$, this gives an element
of the gauge group $G$, so that as $t$ varies from $0$ to $2\pi$, $\Phi[\xi_t]$
traces, as in the abelian case, a closed curve $\Gamma_\Sigma$ in the gauge
group beginning and ending at the identity element.  Depending on the topology
of the gauge group, it may then again happen, as in the abelian theory, that
$\Gamma_\Sigma$ cannot be continuously deformed to a point even when $\Sigma$
itself shrinks to zero.  We have then a topological obstruction inside
$\Sigma$,
which, by analogy to the abelian theory, is designated a monopole.  In
mathematical terms, the monopole charge is defined as the homotopy class of
$\Gamma_\Sigma$ in $G$, which is thus automatically quantised and conserved.

In electromagnetism, a magnetic charge may also be regarded as a source of
the dual Maxwell field $\mbox{\mbox{}$^*\!$} F_{\mu\nu}(x)$ so that its
dynamics can be
formulated in terms of the dual field in exactly the same way that the
dynamics of a electric charge, when considered as a source of $F_{\mu\nu}(x)$,
is formulated in terms of the Maxwell field.  For Yang-Mills theory, however,
for lack of a known dual symmetry, this need no longer be the case.  A colour
magnetic charge as defined above is not a source of the dual Yang-Mills field
$\mbox{\mbox{}$^*\!$} F_{\mu\nu}(x)$ so that its dynamics cannot be
formulated in the same
way as the dynamics of the colour (electric) charge is usually
formulated, in which the latter is taken to be a source of the Yang-Mills
field.

On the other hand, intuitive arguments were given above in the abelian theory
for an intrinsic coupling between a field and its monopole by virtue of the
latter's definition as a topological obstruction of the former.  The same
intuitive argument is applicable also to a monopole in a nonabelian Yang-Mills
field and should lead to a formulation of its dynamics.  Hence, in parallel
to the development of the abelian theory as summarised on the left of Chart II,
we should be able to derive via the same Wu-Yang criterion the equations
of motion for a colour magnetic charge from its definition as a monopole of
the Yang-Mills field.  For that purpose, we ought again to start with the
action of the free field plus the action of a free particle, and impose the
condition that the particle should carry a monopole charge of the Yang-Mills
field.  Extremising the free action then under the defining constraint of
the monopole should automatically lead to coupled equations representing
the interactions between the colour magnetic charge with the Yang-Mills field.
The solution to this problem has already been reported elsewhere.\cite{physrev}
We need here only to summarise the main arguments and results so as to make
clear their relationship with the rest of the paper.

As in the abelian theory, the variational problem posed above is hard to solve
with the gauge potential as field variables since in the presence of a monopole
the gauge potential has to be patched, with the patching dependent on the
position of the monopole.  For the abelian problem, this difficulty was
by-passed by adopting instead the gauge invariant $F_{\mu\nu}(x)$ as field
variables.  Besides avoiding the complexities of patching, $F_{\mu\nu}(x)$
has the added virtue of giving a simple expression for the monopole charge, and
hence a convenient form for the topological constraint we wish to impose.

For the nonabelian theory the field tensor $F_{\mu\nu}(x)$ itself will not do,
but it turns out that the Polyakov loop variable $F_\mu[\xi|s]$ introduced in
Section 3 is eminently suitable for this problem since, in contrast to
$F_{\mu\nu}(x)$, it is patch-independent even in the presence of
a monopole and gives an explicit expression for the monopole charge which
was defined above only abstractly as a homotopy class.

To see this, we recall from Section 3 that $F_\mu[\xi|s]$ may be regarded as
a connection in loop space for which one can construct the curvature
$G_{\mu\nu}[\xi|s,s']$ in (\ref{Gmunu}).  It was further indicated there
that $G_{\mu\nu}[\xi|s]$ represents the phase change under parallel transport
over an infinitesimal rectangle in loop space, which in ordinary space appears
as an infinitesimal version of the surface $\Sigma$ used in Section 2 to
define a monopole.  It is thus clear that $G_{\mu\nu}[\xi|s]$ measures in
some way the monopole charge at the point $\xi(s)$ enclosed by this
infinitesimal surface.  The question in
exactly what way it measures the charge is perhaps not entirely obvious, but
has already been analysed in detail elsewhere. \cite{annphys} For our purpose
here, we need only quote the result for the $SO(3)$ theory where the monopole
charge may be denoted by a sign $\pm$, (i.e. $-$ represents a monopole but
$+$ no monopole).  Then if the surface representing $G_{\mu\nu}[\xi|s]$
encloses a monopole, we have $G_{\mu\nu}[\xi|s] = \kappa$ at $\xi(s)$ but
zero elsewhere, where $\kappa$
satisfies:
\begin{equation}
\exp i\pi \kappa = -1.
\label{expkappa}
\end{equation}

Suppose now we have a classical point particle with mass $m$ moving along a
world-path $Y^\mu(\tau)$.  The defining constraint specifying that
$Y^\mu(\tau)$ should carry a colour magnetic charge can then be written in
a convenient form as:
\begin{equation}
G_{\mu\nu}[\xi|s] = -4\pi J_{\mu\nu}[\xi|s],
\label{Gausslawc}
\end{equation}
where:
\begin{equation}
J_{\mu\nu}[\xi|s] = \tilde{g} \int d\tau \kappa[\xi|s]
   \epsilon_{\mu\nu\rho\sigma} \frac{d Y^\rho(\tau)}{d\tau} \dot{\xi}^\sigma(s)
   \delta(\xi(s)-Y(\tau))
\label{Jmunuc}
\end{equation}
may be regarded as the monopole current, only carrying an extra index here
because in loop space notation, the tangent $\dot{\xi}(s)$ to the loop at
$s$ defines an additional direction.      . (\ref{Gausslawc}) is the
direct nonabelian analogue of the Gauss Law constraint (\ref{gausslawc}) in
the abelian case.  According then to the Wu-Yang criterion,
the equations of motion of the colour magnetic charge are to be obtained by
extremising the free action under this defining constraint.

The free action for the field-particle system is:
\begin{equation}
{\cal A}^0 = {\cal A}_F^0 - m \int d\tau,
\label{ScriptA0c}
\end{equation}
where, in terms of $F_\mu[\xi|s]$ as variables, ${\cal A}_F^0$ takes now the
form (\ref{ScriptA0FL}).  Again, as in the loop formulation of the pure
Yang-Mills theory in Section 3, $F_\mu[\xi|s]$, being a redundant set as field
variables, have to be constrained so as to ensure that the potential $A_\mu(x)$
can be recovered from them.  But, as in the parallel problem for the magnetic
charge in the abelian theory, the beauty of the present formulation is that
the defining constraint (\ref{Gausslawc}) implies that $G_{\mu\nu}[\xi|s]$
vanishes whenever $\xi(s)$ does not lie on the monopole world-line
$Y^\mu(\tau)$
and hence already ensures by the Extended Poincar\'e Lemma that $A_\mu(x)$ can
be recovered everywhere except on $Y^\mu(\tau)$ where the potential is in any
case not expected to exist.  Hence, so long as (\ref{Gausslawc}) is imposed
we are allowed to adopt $F_\mu[\xi|s]$ as variables and just extremise:
\begin{equation}
{\cal A} = {\cal A}^0 + \int \delta \xi ds \,\mbox{\rm Tr} [L_{\mu\nu}[\xi|s]
   \{G^{\mu\nu}[\xi|s] + 4\pi J^{\mu\nu}[\xi|s] \}]
\label{ScriptA}
\end{equation}
with respect to $F_\mu[\xi|s]$ and $Y^\mu(\tau)$, where $L_{\mu\nu}[\xi|s]$
are the Lagrange multipliers for the defining constraint (\ref{Gausslawc}).

Extremising ${\cal A}$ in (\ref{ScriptA}) with respect to $F_\mu[\xi|s]$,
one obtains (\ref{FmuxisinL}) as in the pure field theory leading again to
the Polyakov form (\ref{polyakov}) of the Yang-Mills equation.  Extremising
${\cal A}$ with respect to $Y^\mu(\tau)$, one has:
\begin{eqnarray}
m \frac{d^2 Y^\mu(\tau)}{d \tau^2} & = & -8\pi \tilde{g} \int \delta \xi ds
   \epsilon^{\mu\nu\rho\sigma} \delta^\lambda(s) \,\mbox{\rm Tr}
\{L_{\lambda\rho}[\xi|s]
   \kappa[\xi|s] \} \nonumber \\
   &  & \frac{dY_\nu(\tau)}{d\tau} \dot{\xi}_\sigma(s) \delta(\xi(s)-Y(\tau)),
\label{pardualwong}
\end{eqnarray}
which, together with (\ref{polyakov}) and the constraint (\ref{Gausslawc}),
constitute the equations of motion for the colour magnetic charge interacting
with the Yang-Mills field.

The equation (\ref{pardualwong}) may be written in a more familiar form by
eliminating the Lagrange multiplier $L_{\mu\nu}[\xi|s]$ using the constraint
(\ref{Gausslawc}), giving:
\begin{equation}
m \frac{d^2Y^\mu(\tau)}{d\tau^2} = - \frac{2 \tilde{g}}{\bar{N}}
   \int \delta \xi ds \epsilon^{\mu\nu\rho\sigma} \,\mbox{\rm Tr}
\{\kappa[\xi|s]
   F_\nu[\xi|s] \} \frac{dY_\rho(\tau)}{d\tau} \dot{\xi}_\sigma(s)
   \dot{\xi}(s)^{-2} \delta(\xi(s)-Y(\tau)),
\label{predualwong}
\end{equation}
where on substituting the local expression (\ref{Fmuxisinx}) for
$F_\mu[\xi|s]$,
one has:
\begin{equation}
m \frac{d^2Y^\mu(\tau)}{d\tau^2} = - \tilde{g} \,\mbox{\rm Tr} \{K(\tau)
   \mbox{\mbox{}$^*\!$} F^{\mu\nu}(Y(\tau)) \} \frac{dY_\nu(\tau)}{d\tau},
\label{dualwong}
\end{equation}
with:
\begin{equation}
K(\tau) = \Phi_\xi(s,0) \kappa[\xi|s] \Phi_\xi^{-1}(s,0)|_{\xi(s)=Y(\tau)},
\label{Ktau}
\end{equation}
a local quantity depending only on $Y(\tau)$.  With this, the analogy between
the first column of Charts II and IV is complete.

One notices that (\ref{dualwong}) is very similar in form to the dual Lorentz
equation (\ref{duallorentz}) for the abelian magnetic charge.  It is also
exactly the dual of the equation deduced by Wong \cite{wong} by taking the
classical limit of the Yang-Mills equation and can be obtained from the Wong
equation simply by changing $g \rightarrow {\tilde g}$, $I(\tau) \rightarrow
K(\tau)$, and $F_{\mu\nu}(x) \rightarrow \mbox{\mbox{}$^*\!$} F_{\mu\nu}(x)$.
This does not
mean, however, that the dynamics of a (nonabelian) colour magnetic charge need
be exactly the same as that of a colour (electric) charge as was true for the
abelian case.  The reason is that if so the constraint (\ref{Gausslawc})
would have to be the dual of the second Wong equation:
\begin{equation}
D_\nu F^{\mu\nu}(x) = -4 \pi g \int d\tau I(\tau) \frac{dY^\mu(\tau)}{d\tau}
   \delta(x-Y(\tau))
\label{Wong2}
\end{equation}
but it is not known to be so.

Next, consider a Dirac particle carrying a colour magnetic charge in parallel
to the abelian case as summarised in the second column of Chart II.  As
for the abelian theory we just replace the action (\ref{ScriptA0c})
by:
\begin{equation}
{\cal A}^0 = {\cal A}_F^0 + \int d^4x \bar{\psi}(x) (i \partial_\mu \gamma^\mu
   - m) \psi(x),
\label{ScriptA0q}
\end{equation}
and the classical current (\ref{Jmunuc}) by its Dirac analogue:
\begin{equation}
J_{\mu\nu}[\xi|s] = \tilde{g} \epsilon_{\mu\nu\rho\sigma} [\bar{\psi}(\xi(s))
   \gamma^\rho t^i \psi(\xi(s)))] \Omega_\xi^{-1}(s,0) t_i \Omega_\xi(s,0)
   \dot{\xi}^\sigma(s),
\label{Jmunuq}
\end{equation}
where:
\begin{equation}
\Omega_\xi(s,0) = \omega(\xi(s_+)) \Phi_\xi(s_+,0)
\label{Omegaxis0}
\end{equation}
is a rotation matrix transforming from the colour electric $U$-frame
at the reference point $P_0$ in which $J_{\mu\nu}[\xi|s]$ is measured to the
colour magnetic $\tilde{U}$-frame at $\xi(s)$ in which $\psi(\xi(s))$ is
measured.  The factor $\Phi_\xi(s_+,0)$ transports the $U$-frame at $P_0$
along the loop $\xi$ to the local $U$-frame at $\xi(s)$, while
$\omega(\xi(s))$,
which is a local quantity, rotates the $U$-frame at $x = \xi(s)$ to the
$\tilde{U}$-frame at the same point.  The fact that the values of these
quantities should be taken at $s_+$ will only be of relevance later.

Equations of motion can now be derived by extremising the action
(\ref{ScriptA}) with respect to $F_\mu[\xi|s]$ and $\psi(x)$.  The equation
obtained from varying $F_\mu[\xi|s]$ is the same as before, while that from
varying $\psi(x)$ can be written as:
\begin{equation}
(i \partial_\mu \gamma^\mu - m) \psi(x) = - \tilde{g} \tilde{A}_\mu(x)
   \gamma^\mu \psi(x),
\label{Dualdirac}
\end{equation}
where:
\begin{equation}
\tilde{A}_\mu(x) = - \frac{2 \sqrt{6}}{\bar{N}} \int \delta \xi ds
   \omega(\xi(s_+)) T_{\mu\nu}[\xi|s] \omega^{-1}(\xi(s_+)) \dot{\xi}^\nu(s)
   \dot{\xi}(s)^{-2} \delta(x-\xi(s)).
\label{Atmux}
\end{equation}
The result is again seen to be exactly dual to the corresponding Yang-Mills
equation for colour (electric) charges if one regards $\tilde{A}_\mu(x)$ as
the `dual' of $A_\mu(x)$, but, as noted already for the classical case above,
this need not mean that the dynamics of a colour magnetic and colour electric
charges will be the same since $\tilde{A}_\mu(x)$ is not related to
$\mbox{\mbox{}$^*\!$} F_{\mu\nu}(x)$ in the same manner as $A_\mu(x)$ is
related to
$F_{\mu\nu}(x)$.

In parallel with the last entry in the second column of Chart II for the
abelian
theory, we expect that the present system should also have a ${\tilde U}$-
invariance.  That this is indeed the case is seen if one couples the
transformation (\ref{Uttransf}) for $T_{\mu\nu}[\xi|s]$ with the following
transformation for ${\tilde \psi}(x)$:
\begin{equation}
\psi(x) \longrightarrow [1+i \tilde{g} \tilde{\Lambda}(x)] \psi(x)
\label{Uttransfpsi}
\end{equation}
and for the rotation matrix $\omega(x)$:
\begin{equation}
\omega(x) \longrightarrow [1 +i \tilde{g} \tilde{\Lambda}(x)] \omega(x)
\label{Uttransfomega}
\end{equation}
with:
\begin{equation}
\tilde{\Lambda}(x) = \frac{4 \sqrt{6}}{\bar{N}} \int \delta \xi ds
   \omega(\xi(s_+)) \tilde{\Lambda}_\nu[\xi|s] \omega^{-1}(\xi(s_+))
   \dot{\xi}^\nu(s) \dot{\xi}(s)^{-2} \delta(x-\xi(s)).
\label{Lambdat}
\end{equation}
Now under (\ref{Uttransf}) for $T_{\mu\nu}[\xi|s]$, the pure field terms in
the action ${\cal A}$ of (\ref{ScriptA}) remain invariant, while under the
simultaneous action of (\ref{Uttransf}), (\ref{Uttransfpsi}) and
(\ref{Uttransfomega}), the increment from the second term in (\ref{ScriptA0q})
will cancel the increment from the $\,\mbox{\rm Tr}\{L_{\mu\nu} J^{\mu\nu}\}$
term in
(\ref{ScriptAFL}) yielding thus overall invariance.

The $\tilde{U}$-transformation appears abelian on the dual potential
$T_{\mu\nu}[\xi|s]$, but its action on $\psi(x)$, $\omega(x)$ and $A_\mu(x)$
is nonabelian.  Indeed, by (\ref{Atmux}), one sees that $\tilde{A}_\mu(x)$
transforms as:
\begin{equation}
\tilde{A}_\mu(x) \longrightarrow \tilde{A}_\mu(x) + \partial_\mu
   \tilde{\Lambda}(x) + i \tilde{g} [\tilde{\Lambda}(x), \tilde{A}_\mu(x)],
\label{UttransfAt}
\end{equation}
i.e. exactly as a connection should. Although we have given above explicitly
only the infinitesimal form of the $\tilde{U}$-transformations on $\psi(x)$,
$\omega(x)$ and $\tilde{A}_\mu(x)$, it is clear that by repeating these
transformations (\ref{Uttransfpsi}), (\ref{Uttransfomega}) and
(\ref{UttransfAt}), a $SU(N)$ symmetry will be generated.  The full symmetry
of the field-particle system is thus $SU(N) \times \widetilde{SU(N)}$ as
anticipated, but again, as in the abelian case, since $A_\mu(x)$ and
$\tilde{A}_\mu(x)$ are related, this does not mean that the symmetries
represent two independent degrees of freedom.  The transformation on
$T_{\mu\nu}[\xi|s]$, however, remains abelian in appearance, with
$T_{\mu\nu}[\xi|s]$ forming thus an (unfaithful) abelian representation
of the nonabelian $\widetilde{SU(N)}$ group.

With this observation, the parallel with the second column of Chart II for
the abelian theory is now also complete.

An interesting fact to note is that although $\tilde{A}_\mu(x)$ plays here
the role of parallel phase transport for the wave functions of colour
magnetic charges just as the ordinary Yang-Mills potential $A_\mu(x)$ is the
parallel phase transport for wave functions of colour electric charges,
$\tilde{A}_\mu(x)$ is not known as yet to share the other function of
$A_\mu(x)$ as being the potential from which all other field quantities can
be derived.  Rather, it seems at present that it is the 2-indexed loop
quantity $T_{\mu\nu}[\xi|s]$ which has usurped that function in the dual
formulation.  We shall return to this interesting point briefly in Section 7.

\setcounter{equation}{0}
\section{Yang-Mills Field with Colour (Electric) Charge}

In standard Yang-Mills theory, a Dirac particle carrying a colour (electric)
charge is introduced as a source of the nonabelian gauge field, the dynamics
of which is usually formulated as follows.  We add to the free action
(\ref{ScriptA0q}) an interaction term suggested by the minimal coupling
hypothesis:
\begin{equation}
{\cal A}_I = g \int d^4x \bar{\psi}(x) A_\mu(x) \gamma^\mu \psi(x).
\label{ScriptAIq}
\end{equation}
Then, extremising the total action ${\cal A}_0 + {\cal A}_I$ with respect to
the variables $A_\mu(x)$ and $\psi(x)$, we obtain as equations of motion the
standard Yang-Mills equations:
\begin{equation}
D_\nu F^{\mu\nu}(x) = - 4 \pi g \bar{\psi}(x) \gamma^\mu \psi(x),
\label{Yangmills}
\end{equation}
and:
\begin{equation}
(i \partial_\mu \gamma^\mu - m) \psi(x) = - g A_\mu(x) \gamma^\mu \psi(x),
\label{Dirac}
\end{equation}
where, of course, $F_{\mu\nu}(x)$ is understood to be derivable from $A_\mu(x)$
as per (\ref{Fmunu}).  In the case of a classical point particle, we have
instead as equations of motion the Wong equations, namely (\ref{Wong2}) and:
\begin{equation}
m \frac{d^2 Y^\mu(\tau)}{d\tau^2} = -g \,\mbox{\rm Tr} \{I(\tau)
F^{\mu\nu}(Y(\tau)) \}
   \frac{d Y_\nu(\tau)}{d \tau},
\label{Wong1}
\end{equation}
which were obtained as the classical limit of the corresponding Yang-Mills
equations, namely (\ref{Yangmills}) and (\ref{Dirac}) respectively.
Surprisingly, however, the action corresponding to these Wong equations is
unknown, and it appears that even the concept of an action may have to be
generalised before the Wong equations can be so accommodated in the
conventional approach to the dynamics of a colour (electric) charge.
\cite{yang}

On the other hand, analogy to electromagnetic duality as summarised in Chart II
suggests that there may be an alternative approach to formulating the dynamics
of a colour electric charge moving in a nonabelian gauge field, namely to
consider, if possible, the charge instead as a monopole of the dual field.
In that case, the Wu-Yang criterion can lead to the equations of motion as a
unique consequence by extremising the free action under the defining constraint
of the monopole.  And, if the previous examples are any guide, this should work
both for the classical point particle and for the Dirac particle.  Our purpose
here is to study whether such a procedure actually applies and whether it leads
to the correct equations.

Our first question then is whether a colour electric charge in Yang-Mills
theory
can indeed be considered as a monopole of the `dual field'
$H_{\mu\nu\rho}[\xi|s]$.  It is, however, not entirely clear what a
monopole of $H_{\mu\nu\rho}[\xi|s]$  means since $H_{\mu\nu\rho}[\xi|s]$
is a quantity the likes of which we have not met with before, so that we can
proceed only by analogy to previous examples.  Now in all cases so far
considered, a monopole emerged as a topological obstruction in a construct
as that detailed in Section 2, where one had first
to define a holonomy with respect to some connection giving the parallel
transport for the phases of wave functions.  In the preceding section, for
example, when considering a colour magnetic charge as a monopole of the
Yang-Mills field, the connection employed was the Yang-Mills potential
$A_\mu(x)$
which gave the parallel transport for the phases of the wave functions
$\psi(x)$
of colour electric charges.  So by analogy, now that we are attempting to
treat colour electric charges as monopoles, we would want a connection which
gives the parallel transport for the phases of the wave functions of colour
magnetic charges.  But in the preceding section, we have already met with a
quantity which seems to fulfill this role, namely $\tilde{A}_\mu(x)$ in
(\ref{Atmux}), which transformed under $\tilde{U}$-transformations in
(\ref{UttransfAt}) exactly as a $\tilde{U}$-connection should.  Suppose
that we were to take this $\tilde{A}_\mu(x)$ as connection and construct with
it a phase factor (holonomy) in exactly the same manner as (\ref{Phiofxi}),
then following the procedure of Lubkin \cite{lubkin}, Wu and
Yang \cite{patching}, and Coleman \cite{coleman} as detailed in Section 2,
we should find again topological obstructions which ought by rights to be
regarded as the monopoles dual to the colour magnetic charges.  And if our
conjecture above were indeed correct, then these monopoles in
$\tilde{A}_\mu(x)$ should represent none other than colour electric charges.

On the other hand, colour electric charges are already defined through the
current they carry in, for example, the Wong equation (\ref{Wong2}).  In terms
of the variables $H_{\mu\nu\rho}[\xi|s]$, this equation can be rewritten via
(\ref{Fmuxisinx}) and (\ref{Hmunurho}) as:
\begin{equation}
\delta_\mu \epsilon^{\mu\nu\rho\sigma} H_{\nu\rho\sigma}[\xi|s]
   = -4 \sqrt{6} \pi g \int d\tau I(\tau) \frac{d Y^\mu(\tau)}{d\tau}
   \dot{\xi}_\mu(s) \delta(\xi(s) - Y(\tau)).
\label{YangmillsinH}
\end{equation}
Thus the question of whether a colour electric charge can be considered as a
monopole in $\tilde{A}_\mu(x)$ becomes whether the equation
(\ref{YangmillsinH})
is equivalent to the statement that there is a monopole of $\tilde{A}_\mu(x)$
on the world-line $Y^\mu(\tau)$ in (\ref{YangmillsinH}).

That this is indeed the case can be seen as follows, using arguments closely
echoing those that one would use in the abelian theory to show that an electric
charge is a monopole of $\mbox{\mbox{}$^*\!$} F_{\mu\nu}(x)$ by virtue of the
Maxwell equation
(\ref{maxwelleqc}).  First, multiplying (\ref{YangmillsinH}) by
$\dot{\xi}^\alpha(s) \dot{\xi}(s)^{-2} \delta(x-\xi(s))$ and integrating
over $\delta \xi ds$, one obtains:
\begin{eqnarray}
\int \delta \xi & ds &\delta_\mu(s) \epsilon^{\mu\nu\rho\sigma}
   H_{\nu\rho\sigma}[\xi|s]  \dot{\xi}^\alpha(s) \dot{\xi}(s)^{-2}
   \delta(x-\xi(s)) \nonumber \\
   & = & - \sqrt{6} \pi \bar{N} g \int d\tau I(\tau)
   \frac{d Y^\alpha(\tau)}{d\tau} \delta(x- Y(\tau)).
\label{ProjyminH}
\end{eqnarray}
This equation being Lorentz invariant, we can choose, without loss of
generality, the direction of $Y^\alpha(\tau)$ at $\tau$ to be along the
$0$th direction.  Next, from the fact that space-time is 4-dimensional,
and that the loop quantity $H_{\mu\nu\rho}[\xi|s]$ has by definition
non-vanishing derivatives only in directions transverse to the loop, namely:
\begin{equation}
\delta_\alpha(s) H_{\mu\nu\rho}[\xi|s] \dot{\xi}^\alpha(s) = 0,
\label{transverseH}
\end{equation}
one can show that:
\begin{equation}
\delta^\sigma(s) \epsilon_{\mu\nu\rho\sigma} H^{\mu\nu\rho}[\xi|s]
   \dot{\xi}_0(s) = 3 \delta^i(s) \epsilon_{ijk} H^{\mu jk}[\xi|s]
   \dot{\xi}_\mu(s),
\label{divHid}
\end{equation}
where, as usual, Greek indices run over $0,1,2,3$ but Latin indices run over
only the 3 spatial direction.  We can then rewrite the above equation
(\ref{ProjyminH}) for $\alpha = 0$ as:
\begin{equation}
\partial_i {\cal E}^i(x) = \sqrt{\frac{2}{3}} \pi g \int d\tau I(\tau)
   \delta(x-Y(\tau)),
\label{divcalE}
\end{equation}
with:
\begin{equation}
{\cal E}_i(x) = \frac{1}{\bar{N}} \int \delta \xi ds \epsilon_{ijk}
   H^{\mu jk}[\xi|s] \dot{\xi}_\mu(s) \dot{\xi}(s)^{-2} \delta(x-\xi(s)).
\label{CalE}
\end{equation}

We recall next that since by (\ref{YangmillsinH}), the exterior derivative
of the 3-form $H$ vanishes except on the world-line $Y_\mu(\tau)$ of the
colour electric charge, the Poincar\'e lemma implies that
$H_{\mu\nu\rho}[\xi|s]$ can be expressed in terms of a `dual potential'
$T_{\mu\nu}[\xi|s]$ for any point not on $Y_\mu(\tau)$.  On substituting
then (\ref{HmunurhoinT}) into (\ref{CalE}), one obtains:
\begin{equation}
{\cal E}_i(x) = \epsilon_{ijk} \{ \partial^j {\cal T}^k(x)
   - \partial^k {\cal T}^j(x) \},
\label{CalEinT}
\end{equation}
with:
\begin{equation}
{\cal T}_i(x) = \frac{1}{\bar{N}} \int \delta \xi ds T_{i\mu}[\xi|s]
   \dot{\xi}^\mu(s) \dot{\xi}(s)^{-2} \delta(x-\xi(s)),
\label{CalT}
\end{equation}
where we have used the fact that the loop quantity $T_{\mu\nu}[\xi|s]$
has zero derivative in the direction along the loop.  The Poincar\'e lemma,
however, implies only the local existence of $T_{\mu\nu}[\xi|s]$ and hence also
${\cal T}_i(x)$, but not necessarily their global existence.  Indeed, in view
of the colour electric charge on $Y^\mu(\tau)$, it is clear that ${\cal
T}_i(x)$
must be singular somewhere on any 2-sphere surrounding the colour electric
charge, for if it were not so, then by cutting the sphere into two halves and
applying Stokes' theorem to respectively the northern and southern hemispheres,
we would obtain:
\begin{equation}
\int d^3x \partial^i {\cal E}_i(x) = \oint {\cal E}_i(x) d\sigma^i
   = \oint_+ {\cal T}_i(x) dx^i + \oint_- {\cal T}_i(x) dx^i,
\label{intdivcalE}
\end{equation}
where the 2 resulting line integrals are both taken along the equator but in
opposite directions so that their sum must vanish.  On the other hand, we have
from (\ref{divcalE}) that the integral in (\ref{intdivcalE}) must take the
value $\sqrt{\frac{2}{3}} \pi g I(\tau)$ for any 2-sphere surrounding
$Y(\tau)$.  Hence, we have a contradiction unless ${\cal T}_i$ is singular.

However, this singularity in ${\cal T}_i(x)$, like the Dirac string, is a mere
gauge artifact, and can be avoided by patching.  Introducing then two patches
to cover the 2-sphere under consideration, say for example the northern and
southern hemispheres overlapping along the equator, and defining for each
patch a corresponding ${\cal T}_i^{(N)}(x)$ or ${\cal T}_i^{(S)}$, then within
its own patch, ${\cal T}_i^{(N)}$ or ${\cal T}_i^{(S)}$ need no longer be
singular, but in the overlap region along the equator, they will have to
satisfy the patching condition:
\begin{equation}
\oint_{eq.} \{ {\cal T}_i^{(N)}(x) - {\cal T}_i^{(S)}(x) \} dx^i
   = \sqrt{\frac{2}{3}} \pi g I(\tau),
\label{patchcalT}
\end{equation}
which is the same as (\ref{divcalE}) or equivalent to the Yang-Mills equation
(\ref{YangmillsinH}) defining the colour electric charge.

Our present object then is to show that the condition (\ref{patchcalT}) is
equivalent to the statement that there is an $SO(3)$ monopole of
$\tilde{A}_\mu(x)$ inside the 2-sphere under consideration - i.e. a monopole
defined as before as a topological obstruction via the construct detailed
in Section 2, but with $\tilde{A}_\mu(x)$ as the connection.
As is well-known from the work of Wu and Yang, this latter statement is
the same as saying that $\tilde{A}_\mu(x)$ needs to be patched, and if
we choose to adopt the same two patches as for ${\cal T}_i(x)$ above, then the
`dual potential' $\tilde{A}_\mu(x)$ in respectively the northern and southern
hemispheres will have to satisfy the patching condition:
\begin{equation}
\tilde{A}_\mu^{(S)}(x) - S(x) \tilde{A}_\mu^{(N)}(x) S^{-1}(x)
   = -\frac{i}{\tilde{g}} \partial_\mu S(x) \ S^{-1}(x),
\label{patchAt}
\end{equation}
where $S(x)$, the patching function, has to wind once around the gauge group
$SO(3)$ (i.e. half-way round $SU(2)$) as $x$ winds once around the equator.
Equivalently, we have:
\begin{equation}
P \exp \oint_{eq.} \{ \partial_i S(x) \ S^{-1}(x) \} dx^i = -1.
\label{patchcondS}
\end{equation}
Thus, if we are able to show that the patching condition (\ref{patchcondS})
for $S(x)$ is equivalent to that for ${\cal T}_i(x)$, namely (\ref{patchcalT}),
we will have shown that a colour electric charge is indeed a monopole of
$\tilde{A}_\mu(x)$, and vice versa.  Moreover, since both the conditions
(\ref{patchcalT}) and (\ref{patchcondS}) are invariant under $\tilde{U}$-
transformations (the former because under a $\tilde{U}$-transformation
${\cal T}_i(x)$ changes only by a total derivative whose integral over a closed
loop vanishes), it will be sufficient to demonstrate their equivalence in
some particular $\tilde{U}$-gauge choice.

Now by its definition (\ref{Atmux}) above, one sees that $\tilde{A}_i(x)$
is related to ${\cal T}_i(x)$ by:
\begin{equation}
\omega^{-1}(x) \tilde{A}_i(x) \omega(x) = - 2 \sqrt{6} {\cal T}_i(x),
\label{AtincalT}
\end{equation}
which holds for the patched quantities each in the patch in which it is
defined.
We may thus write:
\begin{equation}
\omega^{(N)-1}(x) \tilde{A}_i^{(N)}(x) \omega^{(N)}(x)
   - \omega^{(S)-1}(x) \tilde{A}_i^{(S)}(x) \omega^{(S)}(x)
   = - 2 \sqrt{6} \{ {\cal T}_i^{(N)}(x) - {\cal T}_i^{(S)}(x) \},
\label{AtincalTNS}
\end{equation}
where we recall that $\omega(x)$ is by definition the rotation matrix which
transforms from the local $U$-gauge to the local $\tilde{U}$-gauge.  Hence,
$\omega^{(S)}(x) \omega^{(N)-1}(x)$ is the matrix which transforms from the
local northern $\tilde{U}$-gauge to the local southern $\tilde{U}$-gauge, or
in other words the patching function is given by:
\begin{equation}
S(x) = \omega^{(S)}(x) \omega^{(N)-1}(x).
\label{Sinomega}
\end{equation}

Thus, from (\ref{AtincalTNS}), we have:
\begin{equation}
\omega^{(S)-1}(x) \{ \tilde{A}_i^{(S)}(x) - S(x) \tilde{A}_i^{(N)}(x)
   S^{-1}(x) \} \omega^{(S)}(x) = 2 \sqrt{6}
   \{ {\cal T}_i^{(N)}(x) - {\cal T}_i^{(S)}(x) \},
\label{AtSincalTNS}
\end{equation}
or by the patching condition (\ref{patchAt}) for $\tilde{A}_i(x)$:
\begin{equation}
\omega^{(S)-1}(x) \{ \partial_i S(x) \ S^{-1}(x) \} \omega^{(S)}(x)
   = 2 \sqrt{6} i \tilde{g} \{ {\cal T}_i^{(N)}(x) - {\cal T}_i^{(S)}(x) \}.
\label{SincalT}
\end{equation}
Next, we go to the special $\tilde{U}$-gauge where both $\omega^{(N)}(x)$
and $\omega^{(S)}(x)$ point in the same fixed `colour' direction.  This is
possible since we can make independent gauge rotations in the two patches.
In that case, $S(x)$ will also point in the same fixed `colour' direction
and commute with $\omega^{(S)}(x)$, giving by (\ref{SincalT}) above:
\begin{equation}
\partial_i S(x) \ S^{-1}(x) = 2 \sqrt{6} i \tilde{g}
   \{ {\cal T}_i^{(N)}(x) - {\cal T}_i^{(S)}(x) \},
\label{SincalTa}
\end{equation}
which shows that in this gauge the right-hand side also points in the same
fixed `colour' direction.  In that case, since the exponents at different
points commute, we can evaluate the holonomy on the left-hand side of
(\ref{patchcondS}) without accounting for the path-ordering, and write just:
\begin{equation}
P \exp \oint_{eq.} \{ \partial_i S(x) \ S^{-1}(x) \} dx^i
   = \exp 2 \sqrt{6} i \tilde{g} \oint_{eq.}
   \{ {\cal T}_i^{(N)}(x) - {\cal T}_i^{(S)}(x)\} dx^i.
\label{holoinoint}
\end{equation}
Recalling then the Dirac quantisation condition relating the unit (colour)
electric and magnetic charges, which for the $su(2)$ theory reads
as: \cite{charges}
\begin{equation}
g \tilde{g} = \frac{1}{4},
\label{su2diracqcon}
\end{equation}
we see that the conditions (\ref{patchcalT}) and (\ref{patchcondS}) are
indeed equivalent as we wanted.

Given now that a colour (electric) charge as usually defined in the Yang-Mills
theory has been shown to be equivalent to a monopole of the dual connection
$\tilde{A}_\mu(x)$, we can then in principle apply the same Wu-Yang criterion
that we have used in all previous cases for monopoles to derive the equations
of motion for a colour electric charge.  All we have to do is to take the free
action of the field-particle system expressed in terms of the
$\tilde{U}$-gauge invariant field variable, namely $H_{\mu\nu\rho}[\xi|s]$, and
extremise with respect to this variable and the particle world-line or wave
function as the case may be, but subject to the defining constraint of the
monopole.  This is a very
different approach to the conventional one for a colour electric charge as
outlined at the beginning of this section, and although the analogy to the
abelian theory suggests that the resultant equations may be the same as the
standard equations, there is {\it a priori} no guarantee at all that this
will indeed be the case.  Nevertheless, as we shall see, the answer does
turn out to be exactly as expected for both a classical and a Dirac point
particle carrying a colour electric charge.

For a classical point charge, the free action is:
\begin{equation}
\tilde{{\cal A}}^0 = \tilde{{\cal A}}_F^0 - m \int d\tau,
\label{ScriptAt0L}
\end{equation}
with $\tilde{{\cal A}}_F^0$ as given in (\ref{ScriptAt0FL}).  The same
arguments as those outlined in Section 5 for the colour magnetic charge
will lead to the conclusion that a colour electric charge in $SO(3)$ theory,
when considered as a monopole of $\tilde{A}_\mu(x)$, can take the values $\pm$,
where $-$ corresponds to a monopole and $+$ no monopole.  Further, the quantity
$\delta_\mu(s) \epsilon^{\mu\nu\rho\sigma} H_{\nu\rho\sigma}[\xi|s]$, which
is the equivalent of $G_{\mu\nu}[\xi|s]$ for the colour magnetic charge, will
take a value $\iota$ satisfying (cf (\ref{expkappa})):
\begin{equation}
\exp i \pi \iota = -1,
\label{expiota}
\end{equation}
at the position of the colour electric charge but is zero elsewhere.  Hence,
in analogy to (\ref{Gausslawc}) and (\ref{Jmunuc}), one can write the
defining constraint for the classical colour electric charge as:
\begin{equation}
\delta_\mu(s) \epsilon^{\mu\nu\rho\sigma} H_{\nu\rho\sigma}[\xi|s]
   = - 4 \pi J[\xi|s]
\label{Dualgauss}
\end{equation}
with:
\begin{equation}
J[\xi|s] = \sqrt{6} g \int d \tau \iota[\xi|s] \frac{dY^\mu(\tau)}{d\tau}
   \dot{\xi}_\mu(s) \delta(\xi(s) - Y(\tau)).
\label{Jxis}
\end{equation}
Extremising then (\ref{ScriptAt0L}) under the constraint (\ref{Dualgauss})
with respect to $H_{\mu\nu\rho}[\xi|s]$ gives the equation (\ref{HmunurhoinL}),
which, as shown in Section 4, is equivalent to the Yang-Mills condition that
the gauge potential $A_\mu(x)$ exists.  Extremising with respect to
$Y^\mu(\tau)$, on the other hand, gives:
\begin{eqnarray}
m \frac{d^2Y^\mu(\tau)}{d\tau^2} & = & 4 \sqrt{6} \pi g \int \delta \xi ds
   \left[\delta_\sigma(s) \,\mbox{\rm Tr} \{ L[\xi|s] \iota[\xi|s] \}
\dot{\xi}^\mu(s)
   \frac{dY^\sigma(\tau)}{d\tau} \right.  \nonumber \\
   & & \left. - \delta^\mu(s) \,\mbox{\rm Tr} \{L[\xi|s] \iota[\xi|s]\}
\dot{\xi}_\sigma(s)
   \frac{dY^\sigma(\tau)}{d\tau} \right] \delta(\xi(s) - Y(\tau)).
\label{protoWong}
\end{eqnarray}
The Lagrange multiplier $L[\xi|s]$ can be eliminated from the equation
(\ref{protoWong}) using the field equation (\ref{HmunurhoinL}) and the fact
that:
\begin{equation}
\delta_\alpha(s) \iota[\xi|s] = 0,
\label{deltaI}
\end{equation}
which is a consequence of the constraint (\ref{YangmillsinH}) and the fact
that $H_{\mu\nu\rho}[\xi|s]$ has zero longitudinal loop derivative.  The
result is:
\begin{eqnarray}
m \frac{d^2Y^\mu(\tau)}{d\tau^2} & = & \sqrt{\frac{2}{3}} \frac{g}{\bar{N}}
   \int \delta \xi ds \left[ \epsilon_{\alpha\nu\rho\sigma}
   \,\mbox{\rm Tr} \{ H^{\alpha\nu\rho}[\xi|s] \iota[\xi|s] \} \dot{\xi}^\mu(s)
   \dot{\xi}(s)^{-2} \frac{dY^\sigma(\tau)}{d\tau} \right. \nonumber \\
   & & - \left. \epsilon^{\alpha\nu\rho\mu} \,\mbox{\rm Tr} \{
H_{\alpha\nu\rho}[\xi|s]
   \iota[\xi|s] \} \dot{\xi}_\sigma(s) \frac{dY^\sigma(\tau)}{d\tau} \right]
   \delta(\xi(s)-Y(\tau)).
\label{proWong}
\end{eqnarray}
By means then of the `duality relation' (\ref{Hmunurho}) giving
$H_{\mu\nu\rho}[\xi|s]$ in terms of $F_\mu[\xi|s]$, one obtains, on
substituting
for $F_\mu[\xi|s]$ its expression (\ref{Fmuxisinx}) in terms of ordinary local
variables, exactly the Wong equation (\ref{Wong1}) in place of (\ref{proWong}).
Similarly, the defining constraint (\ref{YangmillsinH}) itself reduces to the
other Wong equation (\ref{Wong2}).  We have thus, as anticipated, rederived
the Wong equations using an entirely different approach by treating the
colour electric charge as a monopole of the dual connection $\tilde{A}_\mu(x)$
and applying to it the Wu-Yang criterion.  An added interest is that the
equations are now derived from an action principle, which, as we recall, was
not available in the conventional approach for deriving the Wong equations.

Next, for the Dirac particle, we start with the free action:
\begin{equation}
\tilde{{\cal A}}^0 = \tilde{{\cal A}}_F^0 + \int d^4x \bar{\psi}(x)
   (i\partial_\mu \gamma^\mu - m) \psi(x),
\label{ScriptAt0q}
\end{equation}
with $\tilde{{\cal A}}_F^0$ as given in (\ref{ScriptAt0FL}) under the defining
constraint (\ref{YangmillsinH}) of the monopole with:
\begin{equation}
J[\xi|s] = \sqrt{6} g [\bar{\psi}(\xi(s)) t_i \gamma^\mu \psi(\xi(s))]
   t^i \dot{\xi}_\mu(s),
\label{Jxisq}
\end{equation}
where we have again just replaced the classical current in (\ref{Jxis})
by the quantum current.  On extremising:
\begin{equation}
\tilde{{\cal A}} = \tilde{{\cal A}}^0 + \int \delta \xi ds \,\mbox{\rm Tr} \{
L[\xi|s]
   [\delta_\mu(s) \epsilon^{\mu\nu\rho\sigma} H_{\nu\rho\sigma}[\xi|s]
   + 4\pi J[\xi|s]]\}
\label{ScriptAt}
\end{equation}
with respect to $H_{\mu\nu\rho}[\xi|s]$ we obtain the equation
(\ref{HmunurhoinL}), and with respect to $\psi(x)$, the standard Yang-Mills
equation (\ref{Dirac}) for $\psi(x)$ with $A_\mu(x)$ given in terms of the
Lagrange multiplier $L[\xi|s]$ as per (\ref{AmuinW}).  In other words, in as
much as (\ref{HmunurhoinL}) has already been shown in Section 4 to be
equivalent to the usual Yang-Mills field equation with $A_\mu(x)$ as the
gauge potential, we have here rederived exactly the standard Yang-Mills
theory by means of the Wu-Yang criterion treating the colour (electric) charge
as a monopole, and without recourse to either an interaction term in the
action or an appeal to the minimal coupling hypothesis.

Needless to say, these equations, being just the standard Yang-Mills equations,
are invariant under the usual $U$ gauge transformations.  We have thus
completed on the right-hand side of Chart IV the sought-for analogy with the
left-hand side of the same Chart, and also with the right-hand side of Chart
II.

\setcounter{equation}{0}
\section{Remarks}

In the four preceding sections we have shown how, by introducing a generalised
duality relationship for Yang-Mills fields, one can reproduce nonabelian
analogues for all the dual properties of electromagnetism listed in Chart I
and II.  Indeed, the analogy can be made to appear even closer by a few
changes in notation.  Thus, by defining:
\begin{equation}
E_\mu[\xi|s] = \Phi_\xi(s,0) F_\mu[\xi|s] \Phi_\xi^{-1}(s,0),
\label{Emuxis}
\end{equation}
and noting that the constraints signifying the absence of colour electric
and magnetic charges in the pure Yang-Mills theory can be written respectively
as:
\begin{equation}
\delta_\mu E^\mu[\xi|s] = 0,
\label{divEmu}
\end{equation}
and:
\begin{equation}
\delta_\rho H^{\mu\nu\rho}[\xi|s] = 0,
\label{divHmunurho}
\end{equation}
one can recast Chart III into the form shown as Chart V, the analogy of which
to Chart I for pure electromagnetism is very obvious. A similar change of
notation will also reveal a closer analogy of Chart IV with Chart II.

However, the more symmetric-looking Chart V misses one important piece of
information that Chart III contains, namely the fact that the field can
actually be described in terms of just the local potential $A_\mu(x)$, which
is of course essential for the theory to qualify as a local gauge theory.  This
important difference between the charts also highlights a very significant lack
of symmetry between the left- and right-hand sides in each of the charts for
the Yang-mills case, namely between the `direct' and `dual' formulations of
the theory.  Although in both formulations, one has been able to recover local
quantities $A_\mu(x)$ and $\tilde{A}_\mu(x)$ which act as parallel phase
transports for the wave functions of respectively colour electric and colour
magnetic charges, only in the direct formulation has it been shown that the
field can in principle be described just in terms of the local quantity
$A_\mu(x)$.  In other words, the quantity $\tilde{A}_\mu(x)$ in the dual
formulation has acquired so far only the significance of a local phase
transport in parallel to $A_\mu(x)$, but not as yet, like $A_\mu(x)$, also
the significance of a local potential from which all field quantities can
be derived.  As far as we can see at present, it does not seem excluded that
in the dual formulation also, all field quantities may eventually be shown to
be expressible in terms of $\tilde{A}_\mu(x)$, but we have not succeeded in
doing so.  It might even happen that when expressed in terms of
$\tilde{A}_\mu(x)$, an exact dual symmetry would be restored for Yang-Mills
fields through a generalised duality relationship of which we at present
know only the loop space form, namely (\ref{Hmunurho}), but we have no idea
at present whether this may or may not be the case.

In the absence of this knowledge, one can only assume that Yang-Mills theory
is not dual symmetric, or that unlike electromagnetism, (colour) electric and
magnetic charges are governed by different dynamics as regards their
interactions with the field.  For the colour electric charge, one sees that,
in spite of the very different treatment reported above from the conventional,
the dynamics turns out to be exactly the standard local theory of a Yang-Mills
colour source, which is well-known.  For the colour magnetic charge, on the
other hand, we have at present only the `dual' formulation in terms of loop
variables which is unfortunately somewhat unwieldy.  Nevertheless, the
equations
of motion have been derived, as shown above, and with some more persistence,
the formulation can be cast into a Lagrangian form for Feynman diagrams of
colour magnetic charges to be evaluated.  Some initial exploration in this
direction has been made and will be reported elsewhere.\cite{jackie}
  Of course, the
question of whether colour magnetic charges will have any physical significance
can only be answered when more is known about their dynamics.

One novel feature of the above treatment for both (colour) electric and
(colour) magnetic charges is that their dynamics has been derived throughout
using the same universal Wu-Yang criterion for monopoles.  For colour electric
charges, the application of this criterion is made possible by the new result
derived in Section 6 that a colour (electric) charge defined as usual as a
source of the Yang-Mills field turns out to be also a monopole of the dual
phase transport $\tilde{A}_\mu(x)$.  In this treatment, the interaction
between the charge and the field arose purely as the consequence of the
definition of the charge as a topological obstruction in the field.  It gives
a unique answer which coincides exactly with the conventional form of the
interaction deduced from a very different line of argument in all cases where
the latter exist, namely the Lorentz and Dirac equations for ordinary electric
and magnetic charges, and the Wong and Yang-Mills equations for the colour
electric charge.  The result is valid in principle for all Yang-Mills theory
(although we call the field `colour' here for convenience) which means that
all known charges can be interpreted as monopoles and all their physical
forces (except gravitation) have now been derived by this
method.  The approach has arguably some aesthetic appeal versus the
conventional in that monopoles occur `naturally' in gauge theories as
topological obstructions while sources do not seem to have any obvious
geometrical significance, and that their interactions then follow from the
Wu-Yang criterion uniquely without apparently any appeal to the minimal
coupling hypothesis.

For the nonabelian theory, it is interesting to note that a colour charge
(whether electric or magnetic), when considered as a monopole, is by definition
actually labelled only by a homotopy class of the gauge group and has thus
initially no colour direction, i.e. direction in internal symmetry space.
It was in writing down the defining constraint as an equation in the gauge
Lie algebra, namely either (\ref{Gausslawc}), to
formulate the dynamics that we first assign a colour direction to the charge.
It follows then that this direction cannot have any real physical significance,
or that the dynamics must be invariant under both $U$ and $\tilde{U}$ gauge
transformations, as was indeed found to be the case.  Indeed, for the classical
point charge, it is possible to write down the defining constraint as an
equation in the gauge group without giving the charge a colour direction, so
that the dynamics can be at least stated, though not perhaps implemented,
entirely in terms of gauge invariants without introducing a concept of gauge.
However, we do not know what the case is for a quantum charge.

As already stressed in the introduction, the results presented in this paper
concern ordinary Yang-Mills theory in strictly 4 space-time dimensions, only
stated in the somewhat unconventional loop-space language. This loop-space
treatment we believe to be appropriate for the reasons already stated, but
though powerful, the method has certain drawbacks requiring sometimes quite
delicate handling, for example, in having to take limits and derivatives in a
prescribed order which, we think, have to do with the uncertainties in
the functional approach itself.  In this work, we have dealt with the
individual cases as they arose, but have often felt the lack of a general
calculus for handling complex loop space operations.

Finally, it will be interesting to ask in what manner our results above are
related to the recent work of Seiberg, Witten and co-workers.  Given that the
question addressed is in both cases essentially the same, it seems that a
relationship must eventually exist, but since the two parallel approaches
to monopoles have been developed independently for some years, it is not
immediately clear how they are actually related.  However, our result here
may have a complementary value with respect to the other more ambitious
program in that the relationship between the dual field quantities, albeit
rather complicated, is here explicitly given for Yang-Mills fields, a
relationship which, according to Seiberg, is being urgently sought in their
approach.

{\bf Acknowledgement}

One of us (JF) acknowledges the support of the Mihran and Azniv Essefian
Foundation (London), the Soudavar Foundation (Oxford) and the Calouste
Gulbenkian Foundation (Lisbon), while another (TST) thanks the Wingate
Foundation for partial support during most of this work.


\begin{thebibliography}{99}
\bibitem{strom}A. Strominger,
Nucl. Phys. {\bf B343} (1990) 167; C.G. Callan, J.A. Harvey and
A. Strominger, Nucl. Phys. {\bf B359} (1991) 611; {\bf B367} (1991)
60.
\bibitem{duff}M.J. Duff and J.X. Lu, Nucl. Phys. {\bf B354} (1991) 129;
Nucl. Phys. {\bf B357} (1991) 534.
\bibitem{alvarezgaume}E. Alvarez, L. Alvarez-Gaume and Y. Lozano,
Nucl. Phys. {\bf B424} (1994) 155.
\bibitem{gauntlett}J. P. Gauntlett and J. A. Harvey, preprint
EFI-94-36 (hep-th/9407111).
\bibitem{schwarz}A. Sen, Nucl. Phys. {\bf B404} (1993) 109;
Phys. Lett. {\bf B303} (1993) 22; Mod. Phys. Lett. {\bf A8} (1993)
2023; J. H. Schwarz and A. Sen, Nucl. Phys. {\bf B411} (1994) 35.
\bibitem{seibwitt}N. Seiberg and E. Witten, Nucl. Phys. {\bf B426}
(1994) 19; Nucl. Phys. {\bf B431} (1994) 484.
\bibitem{seiberg}K. Intriligator and N. Seiberg, Nucl. Phys. {\bf B431}
(1994) 551; N. Seiberg, Nucl. Phys. {\bf B435} (1995) 129; N. Seiberg,
hepth/9408013, to appear in the Proc. of PASCOS 94.
\bibitem{wittenmon}C. Vafa and E. Witten, Nucl. Phys. {\bf B431} (1994)
3; E. Witten, hep-th/9411102, Math. Res. Lett. {\bf 1} (1994) 769.
\bibitem{wuyang}T. T. Wu and C. N. Yang, Phys. Rev. {\bf D14} (1976)
437.
\bibitem{olivemont}C. Montonen and D. Olive, Phys. Lett. {\bf B72}
(1977) 117.
\bibitem{aharony}O. Aharony, preprint TAUP-2232-95 (hep-th/950213);
L. Girardello, A. Giveon,
M. Porrati and A. Zaffaroni, preprint NYU-TH-94/12/1, IFUM/488/FT,
CPTH-A347-01/95, RI-12-94 (hep-th/9502057).
\bibitem{guyang}C. H. Gu and C. N. Yang, Sci. Sin. {\bf 18} (1975)
483; C. N. Yang, Phys. Rev. {\bf D1} (1970) 2360.
\bibitem{tensorss}M. Henneaux and C. Teitelboim, Phys. Lett. {\bf
B206} (1988) 650; K. Hayashi, Phys. Lett. {\bf 44B} (1973) 497;
M. Kalb and P. Ramond,
Phys. Rev. {\bf D9} (1974) 2273; E. Cremmer and J. Scherk, Nucl. Phys.
{\bf B72} (1974) 117; Y. Nambu, Phys. Reports {\bf 23} (1976) 250.
\bibitem{freedtown}D. Z. Freedman and P. K. Townsend, Nucl. Phys. {\bf
B177} (1981) 282.
\bibitem{wong} S. K. Wong, Nuovo Cim. {\bf 65A} (1970) 689.
\bibitem{jackie}J. Faridani, D. Phil. thesis, Oxford, 1994;
Chan Hong-Mo, J.\ Faridani and Tsou Sheung Tsun, Feynman
Diagrams in Loop space (in preparation).
\bibitem{polyakov}A. M. Polyakov, Nucl. Phys. {\bf B164} (1980) 171.
\bibitem{patching} T. T. Wu and C. N. Yang, Phys. Rev. {\bf D12} (1975)
   3845.
\bibitem{annphys}Chan Hong-Mo, P. Scharbach and Tsou Sheung Tsun,
Ann. Phys. (N.Y.) {\bf 167} (1986) 454; Ann. Phys. (N.Y.) {\bf 166}
(1986) 396.
\bibitem{book}Chan Hong-Mo and Tsou Sheung Tsun, {\it Some Elementary
Gauge Theory Concepts} (World Scientific, 1993).
\bibitem{physrev}Chan Hong-Mo, J. Faridani and Tsou Sheung Tsun,
Phys. Rev. {\bf D51} (1995) 7040.
\bibitem{lubkin} E. Lubkin, Ann. Phys. (N.Y.) {\bf 23} (1963) 233.
\bibitem{coleman} S. Coleman, Erice School (1974) 297.
\bibitem{yang}C.\ N.\ Yang, private communication, 1980.
\bibitem{charges}Chan Hong-Mo and Tsou Sheung Tsun, Nucl. Phys. {\bf
B189} (1981) 364.
\end{thebibliography}
\end{document}